\documentclass[10pt,journal]{IEEEtran}
\usepackage{a0size}
\usepackage{amssymb}
\usepackage{multicol}
\usepackage[english]{babel}
\usepackage{epsfig}
\usepackage{bm}
\usepackage{amsfonts,color,amsmath, lscape,graphics}
\usepackage{color}
\usepackage{url}
\usepackage{epstopdf}
\usepackage{algorithm}
\usepackage{algpseudocode}
\usepackage{cite}
\usepackage{hyperref}
\usepackage{fix2col}
\DeclareMathOperator*{\Maximize}{maximize}
\newtheorem{proposition}{Proposition}
\newtheorem{A_remark}{Remark}
\renewcommand{\figurename}{Fig.}
\addto\captionsenglish{\renewcommand{\figurename}{Fig.}}
\newcommand{\equaltext}[1]{\ensuremath{\stackrel{\text{#1}}{=}}}

   \newcommand{\changet}[1]{{\color{black}#1}}

\begin{document}
\title{{Hybrid Digital and Analog Beamforming Design for Large-Scale Antenna Arrays}
\thanks{Manuscript accepted and to appear in IEEE Journal of Selected Topics in Signal Processing, 2016. This work was supported by the
Natural Sciences and Engineering Research Council (NSERC) of Canada, by Ontario Centres of Excellence (OCE) and by BLiNQ Networks Inc. 
The materials in this paper have been presented in part at IEEE  International Conference on Acoustics, Speech and Signal Processing (ICASSP), Brisbane, Australia, April 2015, and in part at
IEEE International Workshop on Signal Processing Advances in Wireless Communications (SPAWC), Stockholm, Sweden, June 2015.}
\thanks{The authors are with The Edward S. Rogers Sr. Department of Electrical and Computer Engineering, University of Toronto, 10 King's College
Road, Toronto, Ontario M5S 3G4, Canada (e-mails:
\{fsohrabi, weiyu\}@comm.utoronto.ca).}}
\author{Foad~Sohrabi,~\IEEEmembership{Student Member,~IEEE,} and
        Wei~Yu,~\IEEEmembership{Fellow,~IEEE}
}

\maketitle

\begin{abstract}
The potential of using of millimeter wave (mmWave) frequency 
for future wireless cellular communication systems has motivated the study of 
large-scale antenna arrays 
for achieving highly directional beamforming.
 However, the conventional fully digital beamforming methods which require one radio frequency (RF) chain per antenna element is not viable for large-scale antenna arrays due to the high cost and high power consumption of RF chain components in high frequencies. To address the challenge of this hardware limitation, this paper considers a hybrid beamforming architecture in which the overall beamformer consists of a low-dimensional digital beamformer followed by an RF beamformer implemented using analog phase shifters.
 Our aim is to show 
that such an architecture can approach the performance of a fully digital scheme with much fewer number of RF chains. Specifically, this paper establishes that if the number of RF chains is twice the total number of data streams, the hybrid beamforming structure can realize any fully digital beamformer exactly, regardless \changet{of} the number of antenna elements. For cases with fewer number of RF chains, this paper further considers the hybrid beamforming design problem for both the transmission scenario of a point-to-point multiple-input multiple-output (MIMO) system and a downlink multi-user multiple-input single-output (MU-MISO) system. For each scenario, we propose a heuristic hybrid beamforming design that achieves a performance close to the performance of the fully digital beamforming baseline. Finally, the proposed algorithms are modified for the more practical setting in which only finite resolution phase shifters are available. Numerical simulations show that the proposed schemes are effective even when phase shifters with very low resolution are used.
\end{abstract}
\section{Introduction}
Millimeter wave (mmWave) technology is one of the promising candidates for future generation wireless cellular communication systems to address the current challenge of bandwidth shortage \cite{yong2006overview,pi2011introduction,rappaport2013millimeter}. The mmWave signals experience severe path loss, penetration loss and rain fading as compared to signals in current cellular band (3G or LTE) \cite{haider2014cellular}. However, the shorter wavelength at mmWave frequencies
also enables 
 more antennas 
 to be packed
 in the same physical dimension, which allows for large-scale spatial multiplexing and highly directional beamforming.
This leads to the advent of large-scale or massive multiple-input multiple-output (MIMO) concept
for mmWave communications.
Although the principles of the beamforming are the same regardless of carrier frequency, it is not practical to use conventional fully digital beamforming schemes \cite{telatar1999capacity,peel2005vector,wiesel2008zero,dahrouj2010coordinated,shi2011iteratively} for large-scale antenna arrays. This is because the implementation of fully digital beamforming requires one dedicated radio frequency (RF) chain per antenna element,
which is prohibitive from both cost and 
 power consumption perspectives at mmWave frequencies \cite{doan2004design}.

To address the difficulty of limited number of RF chains, this paper considers a two-stage hybrid beamforming architecture in which the beamformer is constructed by concatenation of a low-dimensional digital (baseband) beamformer and an RF (analog) beamformer implemented using phase shifters. In the first part of this paper, we show that 
the number of RF chains in the
hybrid beamforming architecture only needs to scale as twice the total number of data streams
for it to
achieve the exact same performance as that of any fully digital beamforming scheme regardless of the number of antenna elements in the system.

The second part of this paper considers the hybrid beamforming design problem when the number of RF chains is less than twice the number of data streams for two specific scenarios: (i) the point-to-point multiple-input multiple-output (MIMO) communication scenario with large-scale antenna arrays at both ends; (ii)  the downlink multi-user multiple-input single-output (MU-MISO) communication scenario with large-scale antenna array at the base station (BS), but single antenna at each user. For both scenarios, we propose heuristic algorithms to design the hybrid beamformers for the problem of overall spectral efficiency maximization under total power constraint at the transmitter, assuming perfect and instantaneous channel state information (CSI) at the BS and all user terminals. The numerical results suggest that
hybrid beamforming can achieve spectral efficiency close to that of the fully digital solution
with the number of RF chains approximately equal to the number of data streams. Finally, we present a modification of the proposed algorithms for the more practical scenario in which only finite resolution phase shifters are available to construct the RF beamformers.

It should be emphasized that the availability of perfect CSI is an idealistic assumption which rarely occurs in practice, especially for systems implementing large-scale antenna arrays. However, the algorithms proposed in the paper are still useful as a reference point for
studying the performance of hybrid beamforming architecture in comparison with fully digital beamforming. Moreover, for imperfect CSI scenario, one way to design the hybrid beamformers is to first design the RF beamformers assuming perfect CSI, and then to design the digital beamformers employing robust beamforming techniques \cite{Michael_JSTSP,Chalise_2007_CC_DL,caire2010multiuser,Davidson_Foad_ICASSP2013,Ken_outage} to deal with imperfect CSI. It is therefore still of interest to study the RF beamformer design problem in perfect CSI.

To address the challenge of limited number of RF chains, different architectures are studied extensively in the literature. Analog or RF beamforming schemes implemented using analog circuitry are introduced in \cite{venkateswaran2010analog,chen2011multi,tsang2011coding,hur2013millimeter}. They typically use analog phase shifters, which 
impose 
a constant modulus
constraint on the elements of the beamformer. This causes analog beamforming to have poor performance
as compared to the fully digital beamforming designs. Another approach for limiting the number of RF chains is antenna subset selection
which is implemented using simple analog switches \cite{sanayei2004antenna,molisch2005capacity,sudarshan2006channel}. However, they cannot achieve full diversity gain in correlated
channels since only a subset of channels are used in the antenna selection scheme \cite{molisch2003reduced,molisch2004fft}. 

In this paper, we consider the alternative architecture of hybrid digital and analog beamforming which has received significant interest in recent work on large-scale antenna array systems \cite{zhang2005variable,ahmadi2009optimal,el2013spatially,alkhateeb2013hybrid,alkhateeb2014channel,
Wei_Foad_ICASSP2015,endeshaw2014beamforming,chinawei,liang2014low,
Wei_Foad_SPAWC2015,bogale2014hybrid}. The idea of hybrid beamforming is first introduced under the name of antenna soft selection for a point-to-point MIMO scenario \cite{zhang2005variable,ahmadi2009optimal}. It is shown in \cite{zhang2005variable} that for a point-to-point MIMO system with diversity transmission (i.e., the number of data stream is one), hybrid beamforming can realize the optimal fully digital beamformer if and only if the number of RF chains at each end is \changet{at least} two. This paper generalizes the above result for spatial multiplexing transmission for multi-user MIMO systems. In particular, we show that hybrid structure can realize any fully digital beamformer if the number of RF chains is twice the number of data streams. We note that the recent work of \cite{bogale2014hybrid} also addressed the question of how many RF chains are needed for hybrid beamforming structure to realize digital beamforming in frequency selective channels. But, the architecture of hybrid beamforming design used in \cite{bogale2014hybrid} is slightly different from the conventional hybrid beamforming structure in \cite{zhang2005variable,ahmadi2009optimal,el2013spatially,alkhateeb2013hybrid,alkhateeb2014channel,
Wei_Foad_ICASSP2015,endeshaw2014beamforming,chinawei,liang2014low,
Wei_Foad_SPAWC2015}.

The idea of antenna soft selection is reintroduced under the name of hybrid beamforming for mmWave frequencies \cite{el2013spatially,alkhateeb2013hybrid,alkhateeb2014channel}. For a point-to-point large-scale MIMO system, \cite{el2013spatially} proposes an algorithm based on the sparse nature of mmWave channels. It is shown that the spectral efficiency maximization problem for mmWave channels can be approximately solved by minimizing the Frobenius norm of the difference between the optimal fully digital beamformer and the overall hybrid beamformer. Using a compressed sensing algorithm called basis pursuit, \cite{el2013spatially} is able to design the hybrid beamformers which achieve good performance when (i) extremely large number of antennas is used at both ends; (ii) the number of RF chains is strictly greater than the number of data streams; (iii) extremely correlated channel matrix is assumed. But in other cases, there is a significant gap between the 
theoretical maximum
capacity and the achievable rate of the algorithm of \cite{el2013spatially}. This paper devises a heuristic algorithm that reduces this gap for the case that the number of RF chains is equal to the number of data streams; it is also compatible with any channel model.

For the downlink of $K$-user MISO systems, it is shown in \cite{chinawei,liang2014low} that hybrid beamforming with $K$ RF chains at the base station can achieve a reasonable sum rate as compared to the sum rate of fully digital zero-forcing (ZF) beamforming which is near optimal for massive MIMO systems \cite{rusek2013scaling}.
The design of \cite{chinawei,liang2014low} involves matching the 
RF precoder to the phase of the channel and setting the digital precoder to be the ZF beamformer for the effective channel. However, there is still a gap between the rate achieved with this particular hybrid design and the maximum capacity. This paper proposes a method to design hybrid precoders for the case that the number of RF chains is slightly greater than $K$ and numerically shows that the proposed design can be used to reduce the gap to capacity. 

The aforementioned existing hybrid beamforming designs typically
assume the use of infinite resolution phase shifters for implementing analog beamformers. However, the components required for realizing accurate phase shifters can be expensive \cite{montori2010design,krieger2013dense}.
More cost effective low resolution phase shifters are typically used in practice. 
The straightforward way to 
design beamformers with finite resolution phase shifters 
is to design the RF beamformer assuming 
infinite resolution first, 
 then to quantize the value of each phase shifter to a finite set \cite{liang2014low}. However, this approach is not effective for systems with very low resolution phase shifters \cite{Wei_Foad_SPAWC2015}. In the last part of this paper, we present a modification to our proposed method for point-to-point MIMO scenario and multi-user MISO scenario when only finite resolution phase shifters are available. Numerical results in the simulations section show that the proposed method is effective even for the very low resolution phase shifter scenario.

This paper uses capital bold face letters for matrices, small bold face for vectors, and small normal face for scalars. The real part and the imaginary part of a complex scalar $s$ are denoted by $\operatorname{Re}\{s\}$ and $\operatorname{Im}\{s\}$, respectively.  For a column vector $\mathbf{v}$, the element in the $i^\text{th}$ row is denoted by $\mathbf{v}(i)$ while for a matrix $\mathbf{M}$, the element in the $i^\text{th}$ row and the $j^\text{th}$ column is denoted by $\mathbf{M}(i,j)$. Further, we use the superscript ${}^H$ to denote the Hermitian transpose of a matrix
and superscript ${}^*$ to denote the complex conjugate.
 The identity
matrix with appropriate dimensions is denoted by $\mathbf{I}$; $\mathbb{C}^{m\times n}$ denotes an $m$ by $n$ dimensional complex space; $\mathcal{CN}(\mathbf{0},\mathbf{R})$ represents the zero-mean complex Gaussian distribution with covariance matrix $\mathbf{R}$.
Further, the notations $\operatorname{Tr}(\cdot)$, $\operatorname{log}(\cdot)$ and $\mathbb{E}[\cdot] $ represent the
trace, logarithmic and expectation operators, respectively; $|\cdot|$ represent determinant or absolute value depending on context. Finally, 
$\frac{\partial{f}}{\partial{x}}$
is used to denote the partial derivative
of the function $f$ with respect to $x$.

\begin{figure*}[t]
\centering
{\includegraphics[width=0.85\textwidth]{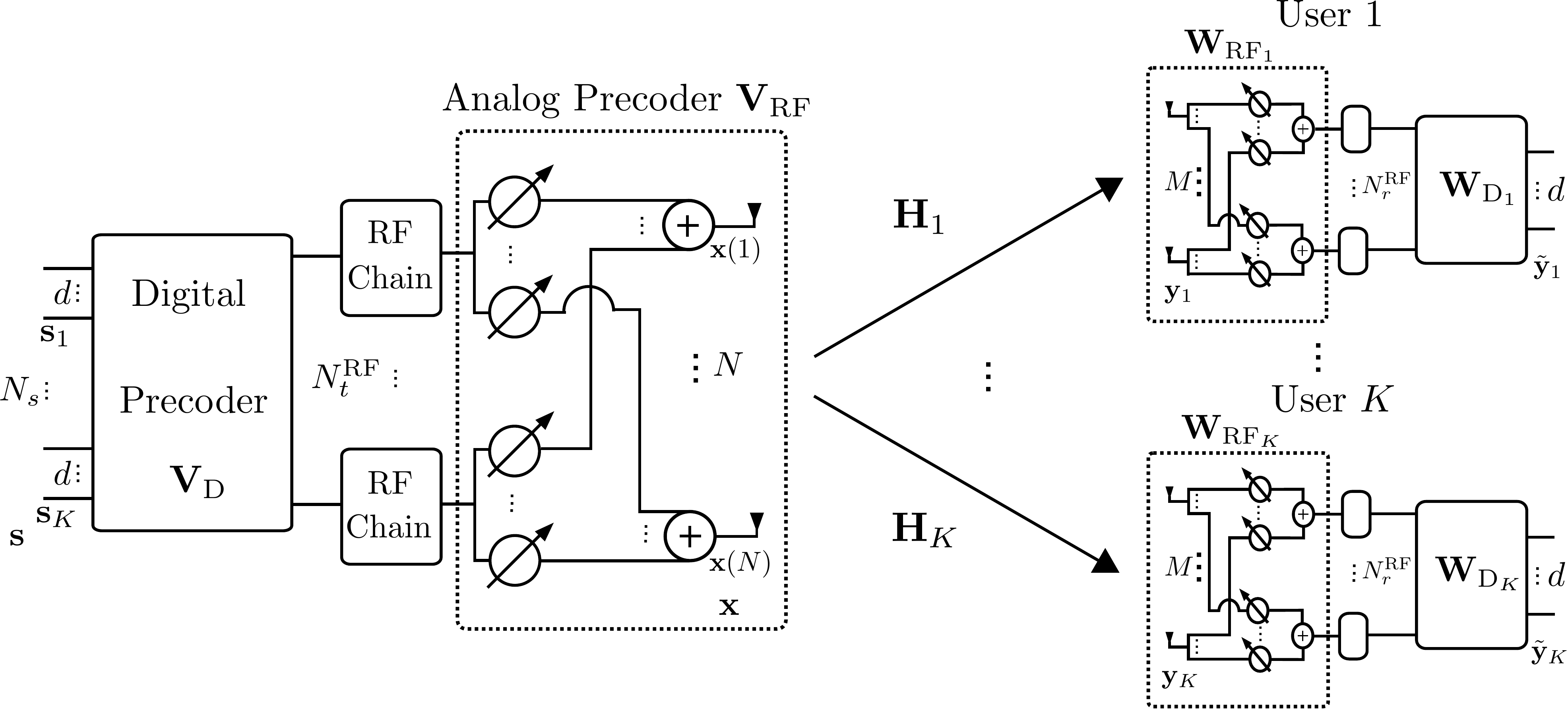}}
\caption{Block diagram of a multi-user MIMO system with hybrid beamforming architecture at the BS and the user terminals.}
\label{fig:sys_model}
\end{figure*}

\section{System Model}
Consider a narrowband downlink single-cell multi-user MIMO system
in which a BS with $N$ antennas and $N_t^\text{RF}$ transmit RF chains serves $K$ users, each equipped with $M$ antennas and $N_r^\text{RF}$ receive RF chains. Further, it is assumed that each user requires $d$ data streams and that $K d \leq N_t^\text{RF} \leq N$ and $d \leq N_r^\text{RF} \leq M$. Since the number of transmit/receive RF chains is limited, the implementation of fully digital beamforming which requires one dedicated RF chain per antenna element, is not possible. Instead, we consider a two-stage hybrid digital and analog beamforming architecture at the BS and the user terminals as shown in Fig.~\ref{fig:sys_model}.

In hybrid beamforming structure, the BS first modifies the data streams digitally at baseband using an $N_t^\text{RF}\times N_s$ digital precoder, $\mathbf{V_\text{D}}$, where $N_s = K d$, then up-converts the processed signals to the carrier frequency by passing through $N_t^\text{RF}$ RF chains. After that, the BS uses an $N\times N_t^\text{RF}$ RF precoder, $\mathbf{V}_\text{RF}$, which is implemented using analog phase shifters, i.e., with $|\mathbf{V}_\text{RF}(i,j) |^2 = 1$, to construct the final transmitted signal. 
Mathematically, the transmitted signal can be written as
\begin{equation}
\mathbf{x} = \mathbf{V}_\text{RF} \mathbf{V}_\text{D} \mathbf{s} =  \sum_{\ell=1}^{K} \mathbf{V_\text{RF}}  \mathbf{V}_{\text{D}_\ell} \mathbf{s}_\ell, 
\end{equation}
where $\mathbf{V}_\text{D} = [\mathbf{V}_{\text{D}_1},\dots,\mathbf{V}_{\text{D}_K}]$, and $\mathbf{s}\in \mathbb{C}^{N_s\times 1}$ is the vector of data symbols which is the concatenation of each user's data stream vector such as $\mathbf{s}= [\mathbf{s}_1^T,\dots,\mathbf{s}_K^T]^T$, where $\mathbf{s}_\ell$ is the data stream vector for user $\ell$. Further, it is assumed that $\mathbb{E}[\mathbf{s}\mathbf{s}^H] = \mathbf{I}_{N_s}$. For user $k$, the received signal can be modeled as
\begin{equation}
\mathbf{y}_k = \mathbf{H}_k \mathbf{V_\text{RF}} \mathbf{V}_{\text{D}_k} \mathbf{s}_k+ \mathbf{H}_k \sum_{\ell\not=k}\mathbf{V_\text{RF}}  \mathbf{V}_{\text{D}_\ell} \mathbf{s}_\ell + \mathbf{z}_k, 
\end{equation}
where $\mathbf{H}_k \in \mathbb{C}^{M\times N}$ is the matrix of complex channel gains from the transmit antennas of the BS to the $k^\text{th}$ user antennas and $\mathbf{z}_k \sim \mathcal{CN}(\mathbf{0},\sigma^2 \mathbf{I}_M) $ denotes additive white Gaussian noise. The user $k$ first processes the received signals using an $M\times N_r^\text{RF}$ RF combiner, $\mathbf{W}_{\text{RF}_{k}}$, implemented using phase shifters such that $|\mathbf{W}_{\text{RF}_k}(i,j)|^2 = 1$, then down-converts the signals to the baseband using $N_r^\text{RF}$ RF chains. Finally, using a low-dimensional digital combiner, $\mathbf{W}_{\text{D}_{k}}\in \mathbb{C}^{N_r^\text{RF}\times d}$, the final processed signals are obtained as
\begin{equation}
\tilde{\mathbf{y}}_k = \underbrace{\mathbf{W}^H_{\text{t}_k} \mathbf{H}_k \mathbf{V}_{\text{t}_k} \mathbf{s}_k}_{\text{desired signals}}+  \underbrace{\mathbf{W}^H_{\text{t}_k} \mathbf{H}_k \sum_{\ell\not=k}\mathbf{V}_{\text{t}_\ell} \mathbf{s}_\ell}_{\text{effective interference}} + \underbrace{\mathbf{W}^H_{\text{t}_k}  \mathbf{z}_k}_{\text{effective noise}},
\end{equation}
where $\mathbf{V}_{\text{t}_k} = \mathbf{V}_\text{RF} \mathbf{V}_{\text{D}_k}$ and $\mathbf{W}_{\text{t}_k} = \mathbf{W}_{\text{RF}_k} \mathbf{W}_{\text{D}_k}$.
 In such a system, the overall spectral efficiency (rate) of user $k$ assuming Gaussian signalling is \changet{\cite{goldsmith2003capacity}}
\begin{equation}
\label{rate}
R_k = \log_2 \Bigl| \mathbf{I}_M + \mathbf{W}_{\text{t}_k}\mathbf{C}_{k}^{-1}\mathbf{W}_{\text{t}_k}^H \mathbf{H}_k \mathbf{V}_{\text{t}_k} \mathbf{V}_{\text{t}_k}^{H} \mathbf{H}_k^{H} \Bigr|,
\end{equation}
where 
$\mathbf{C}_{k}  = \mathbf{W}_{\text{t}_k}^H \mathbf{H}_k \bigl({\sum_{\ell \not= k}} \mathbf{V}_{\text{t}_\ell} \mathbf{V}_{\text{t}_\ell}^{H}\bigr) \mathbf{H}_k^{H} \mathbf{W}_{\text{t}_k} +  \sigma^2 \mathbf{W}_{\text{t}_k}^H \mathbf{W}_{\text{t}_k}$
 is the covariance of the interference plus noise at user $k$.  The problem of interest in this paper is to maximize the overall spectral efficiency under total transmit power constraint, assuming perfect knowledge of $\mathbf{H}_k$, i.e., we aim to find the optimal hybrid precoders at the BS and the optimal hybrid combiners for each user by solving the following problem:
\begin{subequations}
\label{main_problem}
\begin{eqnarray}
&\displaystyle{\Maximize_{\mathbf{V}_\text{RF},\mathbf{V}_\text{D}\mathbf{W}_\text{RF},\mathbf{W}_\text{D}}}  &  \sum_{k=1}^{K} \beta_k R_k\\
&\text{subject to}  &\operatorname{Tr}(\mathbf{V}_\text{RF} \mathbf{V}_\text{D} \mathbf{V}_\text{D}^{H} \mathbf{V}_\text{RF}^{H}) \leq P\\
\label{6c}
&&|\mathbf{V}_\text{RF}(i,j)|^2 = 1, \enspace \forall i,j\\
\label{6d}
&&|\mathbf{W}_{\text{RF}_k}(i,j)|^2 = 1, \enspace \forall i,j,k,
\end{eqnarray}
\end{subequations}
where $P$ is the total power budget at the BS and the weight $\beta_k$ represents the priority of user $k$; \changet{i.e., the larger $\frac{\beta_k}{\sum_{\ell=1}^K \beta_\ell}$ implies greater priority for user $k$}.

The system model in this section is described for a general setting. In the next section, we characterize 
 the minimum number of RF chains in hybrid beamforming architecture for realizing a fully digital beamformer for the general system model. 
 The subsequent parts 
 of the paper focus on two specific scenarios:
\begin{enumerate}
\item Point-to-point MIMO system with large antenna arrays at both ends, i.e., $K=1$ and $\min(N,M) \gg N_s$.
\item Downlink multi-user MISO system with large number of antennas at the BS and single antenna at the user side, i.e., $N\gg K$ and $M=1$.
\end{enumerate}

\section{Minimum Number of RF Chains to Realize Fully Digital Beamformers}
\label{sec:bounds}
The first part of this paper establishes theoretical bounds on the minimum number of 
RF chains that are required for the hybrid beamforming structure to be able to realize any fully digital beamforming schemes. Recall that without the hybrid structure constraints, fully digital beamforming schemes 
can be easily designed with $N_t^\text{RF}=N$ RF chains at the BS and $N_r^\text{RF}=M$ RF chains at the user side  \cite{telatar1999capacity,peel2005vector,wiesel2008zero,dahrouj2010coordinated,
shi2011iteratively}. This section aims to show that hybrid beamforming architecture can realize fully digital beamforming schemes with potentially smaller number of RF chains. 
We begin by presenting a necessary condition on the number of RF chains for implementing a 
 fully digital beamformer, $\mathbf{V}_\text{FD} \in \mathbb{C}^{N\times N_s}$.

\begin{proposition}
\label{lem1} To realize a fully digital beamforming matrix, it is necessary that the number of RF chains in the hybrid architecture \changet{(shown in Fig.~\ref{fig:sys_model})} is greater than or equal to the number of active data streams, i.e., $N^\text{RF} \geq N_s$.
\end{proposition}
\begin{IEEEproof}
It is easy to see that $\operatorname{rank} (\mathbf{V}_\text{RF}\mathbf{V}_\text{D}) \leq N^{\text{RF}}$ and $\operatorname{rank} (\mathbf{V}_\text{FD}) = N_s$. Therefore, hybrid beamforming structure requires at least $N^\text{RF} \geq N_s$ RF chains to implement $\mathbf{V}_\text{FD}$.
\end{IEEEproof}

We now address
how many RF chains are sufficient in the hybrid structure for implementing any fully digital $\mathbf{V}_\text{FD} \in \mathbb{C}^{N\times N_s}$. It is already known that for the case of $N_s = 1$, the hybrid beamforming structure can realize any fully digital beamformer if and only if $N^{\text{RF}}  \geq 2$ \cite{zhang2005variable}. Proposition~\ref{lem2} generalizes this result for any arbitrary value of $N_s$. 
\begin{proposition}
\label{lem2}
To realize any fully digital beamforming matrix, it is sufficient that the number of RF chains in hybrid architecture  \changet{(shown in Fig.~\ref{fig:sys_model})} is greater than or equal to twice the number of data streams, i.e., $N^\text{RF} \geq 2N_s$.
\end{proposition}

\begin{IEEEproof}
Let $N^\text{RF} = 2N_s$ and denote 
$\mathbf{V}_\text{FD} (i,j)=  \changet{\nu_{i,j}} e^{j\phi_{i,j}}$ and $\mathbf{V}_\text{RF}(i,j) = e^{j\theta_{i,j}}$. 
We propose the following solution to satisfy $\mathbf{V}_\text{RF} \mathbf{V}_\text{D} = \mathbf{V}_\text{FD}$. Choose the $k^\text{th}$ column of the digital precoder as $\mathbf{v}^{(k)}_\text{D} = [\mathbf{0}^T \enspace v_{2k-1} \enspace v_{2k} \enspace \mathbf{0}^T]^T$. Then, satisfying $\mathbf{V}_\text{RF} \mathbf{V}_\text{D} = \mathbf{V}_\text{FD}$  is equivalent to
\[
  \left[ {\begin{array}{cccc}
          \dots& e^{j\theta_{i,2k-1}}&e^{j\theta_{i,2k}}&\dots\end{array} } \right]
  \left[ {\begin{array}{c}
         0\\\vdots\\v_{2k-1} \\v_{2k} \\ \vdots \\ 0  \end{array} } \right] = \changet{\nu_{i,j}} e^{j\phi_{i,j}},
\]
or 
\begin{equation}
\label{opt_design_equations}
v_{2k-1} e^{j\theta_{i,2k-1}} + v_{2k} e^{j\theta_{i,2k}} = \changet{\nu_{i,k}} e^{j\phi_{i,k}}, 
\end{equation}
for all $i=1,\dots,N$ and $k=1,\dots,N_s$. This non-linear system of equations has multiple solutions \cite{zhang2005variable}. If we further choose $v_{2k-1}=v_{2k} = \changet{\nu_\text{max}^{(k)}}$ where $\changet{\nu_\text{max}^{(k)} }=\displaystyle{\max_{i}} \{{\changet{\nu_{i,k}}}\}$, it can be verified after several algebraic steps that the following is a solution to \eqref{opt_design_equations}:
\begin{equation} \nonumber
\theta_{i,2k-1} =\phi_{i,k}  - \operatorname{cos}^{-1}\left(\frac{\changet{\nu_{i,k}}}{2\changet{\nu_\text{max}^{(k)}}}\right),
\end{equation}
\begin{equation}
\label{opt_design_equations_PS}
\theta_{i,2k}  = \phi_{i,k}  + \operatorname{cos}^{-1}\left(\frac{\changet{\nu_{i,k}}}{2 \changet{\nu_\text{max}^{(k)}}}\right).
\end{equation} 
Thus for the case that $N^\text{RF} = 2N_s$, a solution to $\mathbf{V}_\text{RF} \mathbf{V}_\text{D} = \mathbf{V}_\text{FD}$ can be readily found.
The validity of the proposition for $N^{\text{RF}} > 2N_s$ is obvious since we can use the same parameters as for $N^{\text{RF}} = 2N_s$ by setting the extra parameters to be zero in $\mathbf{V}_\text{D}$.
\end{IEEEproof}
\begin{A_remark}
\label{Remark_1}
 \normalfont The solution given in Proposition~\ref{lem2} is one possible set of solutions to the equations in \eqref{opt_design_equations}. The interesting property of that specific solution is that as two digital gains of each data stream are identical; i.e., $v_{2k-1}=v_{2k}$, it is possible to convert one realization of the scaled data symbol to RF signal and then use it twice. Therefore, it is in fact  possible to realize any fully digital beamformer using the hybrid structure with $N_s$ RF chains and $2N_sN$ phase shifters. This leads us to the similar result (but with different design) as in \cite{bogale2014hybrid} which considers hybrid beamforming for frequency selective channels. However, in the rest of this paper, we consider the conventional configuration of hybrid structure in which the number of phase shifters are $N^\text{RF}N$. We show that near optimal performance can be obtained with $N^\text{RF}\approx N_s$, thus further reducing the number of phase shifters as compared to the solution above.
\end{A_remark}
\begin{A_remark}
\label{Remark_2}
\normalfont Proposition~\ref{lem2} is stated for the case that $\mathbf{V}_\text{FD}$ is a full-rank matrix, i.e., $\operatorname{rank}(\mathbf{V}_\text{FD}) = N_s$. In the case that $\mathbf{V}_\text{FD}$ is a rank-deficient matrix (which is a common scenario in the low signal-to-noise-ratio (SNR) regime), it can always be decomposed as $\mathbf{V}_\text{FD} = \mathbf{A}_{N\times r}\mathbf{B}_{r\times N_s}$ where $r = \operatorname{rank}(\mathbf{V}_\text{FD})$. Since $\mathbf{A}$ is a full-rank matrix, it can be realized using the procedure in the proof of Proposition~\ref{lem2} as $\mathbf{A}=\mathbf{V}_\text{RF}\mathbf{V}^\prime_\text{D}$ with hybrid structure using $2r$ RF chains. Therefore, $\mathbf{V}_\text{FD} = \mathbf{V}_\text{RF}(\mathbf{V}^\prime_\text{D}\mathbf{B})$ can be realized by hybrid structure using $2r$ RF chains with $\mathbf{V}_\text{RF}$ as RF beamformer and $\mathbf{V}^\prime_\text{D}\mathbf{B}$ as digital beamformer.
\end{A_remark}

\section {Hybrid Beamforming Design for Single-User Large-Scale MIMO Systems}
\label{sec:MIMO}
The second part of this paper considers the design of hybrid beamformers. We first consider a point-to-point large-scale MIMO system in which a BS with $N$ antennas sends $N_s$ data symbols to a user with $M$ antennas where $\min(N,M) \gg N_s$. Without loss of generality, we assume identical number of transmit/receive RF chains, i.e., $N_t^\text{RF}=N_r^\text{RF}=N^\text{RF}$, to simplify the notation. For such a system with hybrid structure, the expression of the spectral efficiency in \eqref{rate} can be simplified to
\begin{equation}
\label{rate_PTP_MIMO}
R = \log_2 \Bigl| \mathbf{I}_M  + \frac{1}{\sigma^2} \mathbf{W}_\text{t} (\mathbf{W}_\text{t}^H \mathbf{W}_\text{t})^{-1}\mathbf{W}_\text{t}^H\mathbf{H} \mathbf{V}_\text{t} \mathbf{V}_\text{t}^{H} \mathbf{H}^{H}\Bigr|.
\end{equation}
where $\mathbf{V}_\text{t}= \mathbf{V}_\text{RF} \mathbf{V}_{\text{D}}$ and $\mathbf{W}_\text{t} = \mathbf{W}_\text{RF} \mathbf{W}_\text{D}$. 

In this section, we first focus on hybrid beamforming design for the case that the number of RF chains  is equal to the number of data streams; i.e., $N^\text{RF}=N_s$. This critical case is important because according to Proposition 1, the hybrid structure requires at least $N_s$ RF chains to be able to realize the fully digital beamformer. For this case, we propose a heuristic algorithm that achieves rate close to capacity.  At the end of this section, we show that by further approximations, the proposed hybrid beamforming design algorithm for $N^\text{RF}=N_s$, can be used for the case of $N_s < N^\text{RF} < 2N_s$ as well.

The problem of rate maximization in \eqref{main_problem} involves joint optimization over the hybrid precoders and combiners. 
\changet{However, the joint transmitter-receive matrix design, for similarly constrained optimization problem is usually found to be difficult to solve \cite{palomar2003joint}. Further, the non-convex constraints on the elements of the analog beamformers in \eqref{6c} and \eqref{6d} make developing low-complexity algorithm for finding the exact optimal solution unlikely \cite{el2013spatially}.}
So, this paper considers the following strategy instead. First, we seek to design the hybrid precoders, assuming that the optimal receiver is used. Then, for the already designed transmitter, we seek to design the hybrid combiner.

The hybrid precoder design problem can be further divided into two steps as follows. The transmitter design problem can be written as
\begin{subequations}
\label{main_problem_transmitter_PTP}
\begin{eqnarray}
\label{obejective_transmitter_PTP}
&\displaystyle{\max_{\mathbf{V}_\text{RF},\mathbf{V}_\text{D}}}  &\log_2 \Bigl| \mathbf{I}_M + \frac{1}{\sigma^2} \mathbf{H} \mathbf{V}_\text{RF} \mathbf{V}_\text{D} \mathbf{V}_\text{D}^{H} \mathbf{V}_\text{RF}^{H} \mathbf{H}^{H}\Bigr| \\
\label{power_constraint_transmitter_PTP}
&\text{s.t.} &\operatorname{Tr}(\mathbf{V}_\text{RF} \mathbf{V}_\text{D}
\mathbf{V}_\text{D}^{H} \mathbf{V}_\text{RF}^{H}) \leq P,\\
\label{unity_const_matrix_PTP}
&&|\mathbf{V}_\text{RF}(i,j)|^2 = 1, \hspace{0.1in}\forall i,j.
\end{eqnarray}
\end{subequations}
This problem is non-convex. This paper proposes the following heuristic algorithm for obtaining a good solution to  \eqref{main_problem_transmitter_PTP}. First, we derive the closed-form solution of the digital precoder in problem \eqref{main_problem_transmitter_PTP} for a fixed RF precoder, $\mathbf{V}_\text{RF}$. It is shown that regardless of the value of $\mathbf{V}_\text{RF}$, the digital precoder typically satisfies $\mathbf{V}_\text{D}\mathbf{V}_\text{D}^H \propto \mathbf{I}$. Then, assuming such a digital precoder, we propose an iterative algorithm to find a local optimal RF precoder.

\subsection{Digital Precoder Design for $N^\text{RF} = N_s$} 
\label{sec:MIMO_Dig}
The first part of the algorithm considers the design of $\mathbf{V}_\text{D}$ assuming that $\mathbf{V}_\text{RF}$ is fixed. For a fixed RF precoder, $\mathbf{H}_\text{eff} = \mathbf{H}\mathbf{V}_\text{RF}$ can be considered as an effective channel and the digital  precoder design problem can be written as
\begin{subequations}
\label{digital_problem_PTP}
\begin{eqnarray}
\label{digital_problem_a_PTP}
&\displaystyle{\max_{\mathbf{V}_\text{D}}} &  \log_2 \bigl| \mathbf{I}_M  + \frac{1}{\sigma^2} \mathbf{H}_\text{eff} \mathbf{V}_\text{D} \mathbf{V}_\text{D}^{H} \mathbf{H}_\text{eff}^{H}\bigr|\\
\label{digital_problem_b_PTP}
&\text{s.t.} & \operatorname{Tr}(\mathbf{Q} \mathbf{V}_\text{D} \mathbf{V}_\text{D}^{H}) \leq P,
\end{eqnarray}
\end{subequations}
where $\mathbf{Q} = \mathbf{V}_\text{RF}^H \mathbf{V}_\text{RF}$. This problem has a well-known water-filling solution as 
\begin{equation}
\label{optimal_digital_precode_PTP}
\mathbf{V}_\text{D} = \mathbf{Q}^{-1/2} \mathbf{U}_e \boldsymbol{\Gamma}_e,
\end{equation}
 where $\mathbf{U}_e$ is the set of right singular vectors corresponding to the $N_s$ largest singular values of $ \mathbf{H}_\text{eff} \mathbf{Q}^{-1/2}$ and $\boldsymbol{\Gamma}_e$ is the diagonal matrix of allocated powers to each stream.  

Note that for large-scale MIMO systems, $\mathbf{Q} \approx N\mathbf{I}$ with high probability \cite{el2013spatially}. This is because the diagonal elements of $\mathbf{Q}=\mathbf{V}_\text{RF}^H \mathbf{V}_\text{RF} $ are exactly $N$ while the off-diagonal elements can be approximated as a summation of $N$ independent terms which is much less than $N$ with high probability for large $N$. This property enables us to show that the optimal digital precoder for $N^\text{RF}=N_s$ typically satisfies $\mathbf{V}_\text{D}\mathbf{V}_\text{D}^H \propto \mathbf{I}$. The proportionality constant can be obtained with further assumption of equal power allocation for all streams, i.e., $\boldsymbol{\Gamma}_e \approx \sqrt{{{P}}/{{N^\text{RF}}}} \mathbf{I}$. So, optimal digital precoder is $\mathbf{V}_\text{D} \approx \gamma \mathbf{U}_e$ where $\gamma^2 ={{{P}}/({{N N^\text{RF}}})}$. Since $\mathbf{U}_e$ is a unitary matrix for the case that $N^\text{RF} = N_s$, we have $ \mathbf{V}_\text{D}   \mathbf{V}_\text{D}^H \approx \gamma^2 \mathbf{I}$.

\subsection{RF Precoder Design for $N^\text{RF} = N_s$}
Now, we seek to design the RF precoder assuming $ \mathbf{V}_\text{D}   \mathbf{V}_\text{D}^H \approx \gamma^2 \mathbf{I}$. Under this assumption, the transmitter power constraint \eqref{power_constraint_transmitter_PTP} is automatically satisfied for any design of $\mathbf{V}_\text{RF}$. Therefore, the RF precoder can be obtained by solving
\begin{subequations}
\label{main_problem_transmitter_RF_PTP}
\begin{eqnarray}
\label{obejective_transmitter_RF_PTP}
 &\displaystyle{\max_{\mathbf{V}_\text{RF}}}  &\log_2 \Bigl| \mathbf{I}  + \frac{\gamma^2}{\sigma^2} \mathbf{V}_\text{RF}^{H} \mathbf{F}_1 \mathbf{V}_\text{RF} \Bigr|\\
\label{unity_const_matrix_RF_PTP}
&\text{s.t.} &|\mathbf{V}_\text{RF}(i,j)|^2 = 1, \hspace{0.1in}\forall i,j,
\end{eqnarray}
\end{subequations}
where  
$\mathbf{F}_1 = \mathbf{H}^H \mathbf{H}$.
This problem is still non-convex, since the objective function of \eqref{main_problem_transmitter_RF_PTP} is not concave in $\mathbf{V}_\text{RF}$. However, the decoupled nature of the constraints in this formulation enables us to devise an iterative coordinate descent algorithm over the elements of the RF precoder.

\changet{In order to extract the contribution of ${\mathbf{V}_\text{RF}(i,j)}$ to the objective function of \eqref{main_problem_transmitter_RF_PTP}, it is shown in \cite{Wei_Foad_SPAWC2015,pi2012optimal} that the objective function in \eqref{main_problem_transmitter_RF_PTP} can be rewritten as }
\begin{equation}
\label{isolation}
\log_2 \bigl| \mathbf{C}_j \bigr| + \log_2 \left( 2\operatorname{Re}\bigl\{ {{\mathbf{V}_\text{RF}^{*}}(i,j)} \eta_{ij} \bigr\} + \zeta_{ij} +1\right),
\end{equation}
where
\begin{eqnarray*} 
 \mathbf{C}_j = \mathbf{I} + \frac{\gamma^2}{\sigma^2} { 
({{\bar{\mathbf{V}}}_\text{RF}^j})^H   
\mathbf{F}_1 {\bar{\mathbf{V}}_\text{RF}^{j}}},
\end{eqnarray*}
and 
${\bar{\mathbf{V}}_\text{RF}^{j}}$
  is the sub-matrix of $\mathbf{V}_\text{RF}$ with $j^\text{th}$ column removed,
\begin{eqnarray*} 
\eta_{ij} &=& { \displaystyle{\sum_{\ell\not=i}} } \mathbf{G}_j(i,\ell) {\mathbf{V}_\text{RF}(\ell,j)}, \\
\zeta_{ij} &=&  \mathbf{G}_j(i,i) \\
&& +2\operatorname{Re} \left\{ \displaystyle{\sum_{m\not=i, n\not=i}}{{{\mathbf{V}_\text{RF}^{*}}(m,j)}}\mathbf{G}_j(m,n) {\mathbf{V}_\text{RF}(n,j)}  \right\} , 
\end{eqnarray*} 
and 
$\mathbf{G}_j =  \frac{\gamma^2}{\sigma^2} \mathbf{F}_1 - \frac{\gamma^4}{\sigma^4} \mathbf{F}_1   {\bar{\mathbf{V}}_\text{RF}^{j}}  \mathbf{C}_j^{-1} ({\bar{\mathbf{V}}_\text{RF}^{j}})^H   \mathbf{F}_1 $. 
  Since $\mathbf{C}_j$, $\zeta_{ij}$ and $\eta_{ij}$ are independent of $\mathbf{V}_\text{RF}(i,j)$, if we assume that all elements of the RF precoder are fixed except $\mathbf{V}_\text{RF}(i,j)$, the optimal value for the element of the RF precoder at the $i^\text{th}$ row and $j^\text{th}$ column is given by 
\begin{align}
\label{sequential_update_PTP}
\mathbf{V}_\text{RF}(i,j) =   
\begin{cases} 1, &\mbox{if } \eta_{ij}=0, \\
\frac{\eta_{ij}}{|\eta_{ij}|}, & \mbox{otherwise}. \end{cases}
\end{align}

This enables us to propose an iterative algorithm that starts with an initial feasible RF precoder satisfying \eqref{unity_const_matrix_RF_PTP}, i.e., $\mathbf{V}_\text{RF}^{(0)} = \mathbf{1}_{N\times N^\text{RF}}$, then sequentially updates each element of RF precoder according to \eqref{sequential_update_PTP} until the algorithm converges to a local optimal solution of $\mathbf{V}_\text{RF}$ of the problem \eqref{main_problem_transmitter_RF_PTP}. \changet{Note that since in each element update step of the proposed algorithm, the objective function of \eqref{main_problem_transmitter_RF_PTP} increases (or at least does not decrease), therefore the convergence of the algorithm is guaranteed.} The proposed algorithm for designing the RF beamformer in \eqref{main_problem_transmitter_RF_PTP} is summarized in Algorithm~\ref{Alg:1}.
We mention that the proposed algorithm is inspired by the algorithm in \cite{pi2012optimal} that seeks to solve the problem 
of transmitter precoder design with per-antenna power constraint which happens to have the same form as the problem in \eqref{main_problem_transmitter_RF_PTP}.

\begin{algorithm}[t]
\caption{Design of $\mathbf{V}_{\text{RF}}$ by solving \eqref{main_problem_transmitter_RF_PTP}}
\label{Alg:1}
\textbf{Require:} $\mathbf{F}_1$, $\gamma^{2}$, $\sigma^2$
\begin{algorithmic}[1]
\State Initialize $\mathbf{V}_\text{RF} = \mathbf{1}_{N\times N^\text{RF}}$.
		   	\For{$j= 1 \to N^\text{RF}$}\label{step}
		\State Calculate $\mathbf{C}_j = \mathbf{I} + \frac{\gamma^2}{\sigma^2} { 
		({{\bar{\mathbf{V}}}_\text{RF}^j})^H \mathbf{F}_1 {\bar{\mathbf{V}}_\text{RF}^{j}}}$.
		\State Calculate $\mathbf{G}_j =  \frac{\gamma^2}{\sigma^2} \mathbf{F}_1 - \frac{\gamma^4}{\sigma^4} 				\mathbf{F}_1   {\bar{\mathbf{V}}_\text{RF}^{j}}  \mathbf{C}_j^{-1} 									({\bar{\mathbf{V}}_\text{RF}^{j}})^H   \mathbf{F}_1 $.
         		\For{$i= 1 \to N$}
         		\State Find $\eta_{ij} = { \sum_{\ell\not=i}} \mathbf{G}_j(i,\ell) {\mathbf{V}_\text{RF}(\ell,j)}$.
         		\State $\mathbf{V}_\text{RF}(i,j) =   
\begin{cases} 1, &\mbox{if } \eta_{ij}=0, \\
\frac{\eta_{ij}}{|\eta_{ij}|}, & \mbox{otherwise}. \end{cases}$
         		\EndFor
          \EndFor
         	\State Check convergence. If yes, stop; if not go to Step \ref{step}.
\end{algorithmic}
\end{algorithm}

\subsection{Hybrid Combining Design for $N^\text{RF} = N_s$} 
Finally, we seek to design the hybrid combiners 
that maximize the overall spectral efficiency in \eqref{rate_PTP_MIMO} assuming that the hybrid precoders are already designed. For the case that $N^\text{RF}=N_s$, the digital combiner is a square matrix with no constraint on its entries. Therefore, without loss of optimality, the design of $\mathbf{W}_\text{RF}$ and $\mathbf{W}_\text{D}$ can be decoupled by first designing the RF combiner assuming optimal digital combiner and then finding the optimal digital combiner for that RF combiner. As a result, the RF combiner design problem can be written as 
\begin{subequations}
\label{problem_receiver_TF_PTP}
\begin{eqnarray}
&\displaystyle{\max_{\mathbf{W}_\text{RF}}}  &\log_2 \Bigl| \mathbf{I}  + \frac{1}{\sigma^2} (\mathbf{W}_\text{RF}^H \mathbf{W}_\text{RF})^{-1}\mathbf{W}_\text{RF}^H \mathbf{F}_2 \mathbf{W}_\text{RF}\Bigr|  \\
&\text{s.t.} &|\mathbf{W}_\text{RF}(i,j)|^2 = 1, \hspace{0.1in}\forall i,j,
\end{eqnarray}
\end{subequations}
where $\mathbf{F}_2 = \mathbf{H} \mathbf{V}_\text{t} \mathbf{V}_\text{t}^{H} \mathbf{H}^{H}$. This problem is very similar to the RF precoder design problem in \eqref{main_problem_transmitter_RF_PTP}, except the extra term  $(\mathbf{W}_\text{RF}^H\mathbf{W}_\text{RF})^{-1}$. 
Analogous to 
the argument made in Section~\ref{sec:MIMO_Dig} for 
the RF precoder, it can be shown that the RF combiner typically satisfies $\mathbf{W}_\text{RF}^H\mathbf{W}_\text{RF} \approx M\mathbf{I}$, for large $M$. Therefore, the problem \eqref{problem_receiver_TF_PTP} can be approximated in the form of RF precoder design problem in \eqref{main_problem_transmitter_RF_PTP} and Algorithm~\ref{Alg:1} can be used to design $\mathbf{W}_\text{RF}$ by substituting $\mathbf{F}_2$ and $\frac{1}{M}$ by $\mathbf{F}_1$ and $\gamma^2$, respectively, i.e., 
\begin{subequations}
\label{main_problem_app_receiver_RF_PTP}
\begin{eqnarray}
 &\displaystyle{\max_{\mathbf{W}_\text{RF}}}  &\log_2 \Bigl| \mathbf{I}  + \frac{1}{M\sigma^2} \mathbf{W}_\text{RF}^{H} \mathbf{F}_2 \mathbf{W}_\text{RF} \Bigr|\\
&\text{s.t.} &|\mathbf{W}_\text{RF}(i,j)|^2 = 1, \hspace{0.1in}\forall i,j.
\end{eqnarray}
\end{subequations}

Finally, assuming all other beamformers are fixed, the optimal digital combiner is the MMSE solution as
\begin{equation}
\mathbf{W}_\text{D} = \mathbf{J}^{-1} \mathbf{W}_\text{RF}^H \mathbf{H} \mathbf{V}_\text{t},
\end{equation}
where $\mathbf{J} = \mathbf{W}_\text{RF}^H \mathbf{H} \mathbf{V}_\text{t} \mathbf{V}_\text{t}^H \mathbf{H}^H \mathbf{W}_\text{RF} + \sigma^2 \mathbf{W}_\text{RF}^H \mathbf{W}_\text{RF} $.

\subsection{Hybrid Beamforming Design for $N_s<N^\text{RF} < 2N_s$}
\label{subsec:between_extremes}
In Section~\ref{sec:bounds}, we show how to design the hybrid beamformers for the case $N^\text{RF} \geq 2N_s$ for which the hybrid structure can achieve the same rate as the rate of optimal fully digital beamforming. Earlier in this section, we propose a heuristic hybrid beamforming design algorithm for $N^\text{RF} =N_s$. Now, we aim to design the hybrid beamformers for the case of $N_s<N^\text{RF}<2N_s$.

For $N_s<N^\text{RF}<2N_s$, the transmitter design problem can still be formulated as in \eqref{main_problem_transmitter_PTP}. For a fixed RF precoder, it can be seen that the optimal digital precoder can still be found according to \eqref{optimal_digital_precode_PTP}, however now it satisfies $\mathbf{V}_\text{D}\mathbf{V}_\text{D}^H \approx \gamma^2 [\mathbf{I}_{N_s} \enspace \mathbf{0}]$. For such a digital precoder, the objective function of \eqref{main_problem_transmitter_PTP} that should be maximized over $\mathbf{V}_\text{RF}$ can be rewritten as
\begin{equation} 
\label{eigenvalues_RF_design}
\log_2 {\prod_{i=1}^{N_s}} \left(1+\frac{\gamma^2}{\sigma^2} \lambda_i\right),
\end{equation}
where $\lambda_i$ is the $i^\text{th}$ largest eigenvalues of $\mathbf{V}_\text{RF}^{H} \mathbf{H}^{H} \mathbf{H} \mathbf{V}_\text{RF}$. Due to the difficulties of optimizing over a function of subset of eigenvalues of a matrix, we approximate \eqref{eigenvalues_RF_design} with an expression including all of the eigenvalues, i.e., 
$\log_2{\prod_{i=1}^{N^\text{RF}}} (1+\frac{\gamma^2}{\sigma^2} \lambda_i) ,$
or equivalently,
\begin{equation}
\log_2 \Bigl| \mathbf{I}_{N^\text{RF}}  + \frac{\gamma^2}{\sigma^2} \mathbf{V}_\text{RF}^{H} \mathbf{H}^{H} \mathbf{H} \mathbf{V}_\text{RF} \Bigr|,
\end{equation}
which is a reasonable approximation for the practical settings where $N^\text{RF}$ is in the order of $N_s$. Further, by this approximation, the RF precoder design problem is now in the form of \eqref{main_problem_transmitter_RF_PTP}. Hence, Algorithm~\ref{Alg:1} can be used to obtain the RF precoder. In summary, we suggest to first design the RF precoder assuming that the number of data streams is equal to the number of RF chains, then for that RF precoder, to obtain the digital precoder for the actual $N_s$.

At the receiver, we still suggest to design the RF combiner first, then set the digital combiner to the MMSE solution. This decoupled optimization of RF combiner and digital combiner is approximately optimal for the following reason. Assume that all the beamformers are already designed except the digital combiner. Since $\mathbf{W}_\text{RF}^H \mathbf{W}_\text{RF}\approx M\mathbf{I}$, the effective noise after the RF combiner can be considered as an uncolored noise with covariance matrix $\sigma^2 M\mathbf{I}$. Under this condition, by choosing the digital combiner as the MMSE solution, the mutual information between the data symbols and the processed signals before digital combiner is approximately equal to the mutual information between the data symbols and the final processed signals. Therefore, it is approximately optimal to first design the RF combiner using Algorithm~\ref{Alg:1}, then set the digital combiner to the MMSE solution.

\begin{algorithm}[t]
\caption{Design of Hybrid Beamformers for Point-to-Point MIMO systems}
\label{Alg:2}
\textbf{Require:} $\sigma^2$, $P$
\begin{algorithmic}[1]
	\State Assuming $\mathbf{V}_\text{D} \mathbf{V}_\text{D}^H = \gamma \mathbf{I}$ where $\gamma = \sqrt{{{{P}}/({{N N^\text{RF}}}})}$, find $\mathbf{V}_\text{RF}$ by solving the problem in 
	\changet{\eqref{main_problem_transmitter_RF_PTP}}
	using Algorithm~\ref{Alg:1}.
	\State Calculate $\mathbf{V}_\text{D} = (\mathbf{V}_\text{RF}^H \mathbf{V}_\text{RF})^{-1/2} \mathbf{U}_e \boldsymbol{\Gamma}_e$ where $\mathbf{U}_e$ and $\boldsymbol{\Gamma}_e$ are defined as following \eqref{optimal_digital_precode_PTP}.
	\State Find $\mathbf{W}_\text{RF}$ by solving the problem in \eqref{main_problem_app_receiver_RF_PTP} using Algorithm~\ref{Alg:1}.
	\State Calculate $\mathbf{W}_\text{D} = \mathbf{J}^{-1} \mathbf{W}_\text{RF}^H \mathbf{H} \mathbf{V}_\text{RF}\mathbf{V}_\text{D}$ where $\mathbf{J} = \mathbf{W}_\text{RF}^H \mathbf{H} \mathbf{V}_\text{RF}\mathbf{V}_\text{D} \mathbf{V}_\text{D}^H \mathbf{V}_\text{RF}^H \mathbf{H}^H \mathbf{W}_\text{RF} + \sigma^2 \mathbf{W}_\text{RF}^H \mathbf{W}_\text{RF} $.
\end{algorithmic}
\end{algorithm}

The summary of the overall proposed procedure for designing the hybrid beamformers for spectral efficiency maximization in a large-scale point-to-point MIMO system is given in Algorithm~\ref{Alg:2}. \changet{Assuming the number of antennas at both ends are in the same range, i.e., $M=O(N)$, it can be shown that the overall complexity of Algorithm~\ref{Alg:2} is $O(N^3)$ which is similar to the most of the existing hybrid beamforming designs, i.e., the hybrid beamforming designs in \cite{zhang2005variable,el2013spatially}.} 

Numerical results presented in the simulation part of this paper suggest that for the case of $N^\text{RF}=N_s$ and infinite resolution phase shifters, the achievable rate of the proposed algorithm is very close the maximum capacity. The case of $N_s < N^{RF} < 2N_s$ is of most interest when the finite resolution phase shifters are used. It is shown in the simulation part of this paper that the extra number of RF chains can be used to trade off the accuracy of the phase shifters.

\section {Hybrid Beamforming Design for Multi-User Massive MISO Systems}
\label{sec:MISO}

Now, we consider the design of hybrid precoders for the downlink MU-MISO system in which a BS with large number of antennas $N$, but limited number of RF chains $N^\text{RF}$, supports $K$ single-antenna users where $N\gg K$. For such a system with hybrid precoding architecture at the BS, the rate expression for user $k$ in \eqref{rate} can be expressed as
\begin{equation}
R_k = \log_2 \left( 1 + \frac{|\mathbf{h}_k^H \mathbf{V}_\text{RF} \mathbf{v}_{\text{D}_k}|^2 } {\sigma^2 + \sum_{\ell \not= k}  |\mathbf{h}_k^H \mathbf{V}_\text{RF} \mathbf{v}_{\text{D}_\ell}|^2 } \right),
\end{equation}
where $\mathbf{h}_k^H$ is the channel from the BS to the $k^\text{th}$ user and $\mathbf{v}_{\text{D}_\ell}$ denotes the $\ell^\text{th}$ column of the digital precoder $\mathbf{V}_\text{D}$. The problem of overall spectral efficiency maximization for the MU-MISO systems differs from that for the point-to-point MIMO systems in two respects. First, in the MU-MISO case the receiving antennas are not collocated, therefore 
we cannot use the rate expression in \eqref{rate_PTP_MIMO}, which assumes cooperation between the receivers.
The hybrid beamforming design for MU-MISO systems 
must account for the effect of inter-user interference. Second, the priority of the streams may be unequal in a MU-MISO system,
while different streams in a point-to-point MIMO systems always have the same priority. This section considers the hybrid beaforming design of a MU-MISO system to maximize the weighted sum rate.

In \cite{chinawei,liang2014low}, it is shown for the case $N^\text{RF}=K$ and $N \to \infty$, that by matching the RF precoder to the overall channel (or the strongest paths of the channel) and using a low-dimensional zero-forcing (ZF) digital precoder, the hybrid beamforming structure can achieve a reasonable sum rate as compared to the sum rate of fully digital ZF scheme (which is near optimal in massive MIMO systems \cite{rusek2013scaling}). However, for practical values of $N$, there is still a gap between the achievable rates and the capacity. This section proposes a design for the scenarios where $N^\text{RF}> K$ with practical $N$ and show numerically that 
adding a few more RF chains can increase the overall performance of the system and reduce the gap to capacity.

Solving the problem \eqref{main_problem} for such a system involves a joint optimization over $\mathbf{V}_\text{RF}$ and $\mathbf{V}_\text{D}$ which is challenging. We again decouple the design of $\mathbf{V}_\text{RF}$ and $\mathbf{V}_\text{D}$ by considering ZF beamforming with power allocation as the digital precoder. We show that the optimal digital precoder with such a structure can be found for a fixed RF precoder. In addition, for a fixed power allocation, an approximately local-optimal RF precoder can be obtained. By iterating between those designs, a good solution of the problem \eqref{main_problem} for MU-MISO can be found.
\subsection{Digital Precoder Design}
We consider ZF beamforming with power allocation as the low-dimensional digital precoder part of the BS's precoder to manage the inter-user interference. For a fixed RF precoder, such a digital precoder can be found as \cite{peel2005vector}
\begin{equation}
\label{ZF_structure}
\mathbf{V}_\text{D}^\text{ZF} = \mathbf{V}_\text{RF}^H \mathbf{H}^H (\mathbf{H}  \mathbf{V}_\text{RF} \mathbf{V}_\text{RF}^H    \mathbf{H}^H)^{-1} \mathbf{P}^{\frac{1}{2}} = \tilde{\mathbf{V}}_\text{D}  \mathbf{P}^{\frac{1}{2}}, 
\end{equation}
where $\mathbf{H} = [\mathbf{h}_1,\dots,\mathbf{h}_K]^{H}$, $\tilde{\mathbf{V}}_\text{D} =  \mathbf{V}_\text{RF}^H \mathbf{H}^H (\mathbf{H}  \mathbf{V}_\text{RF} \mathbf{V}_\text{RF}^H    \mathbf{H}^H)^{-1}$ and $\mathbf{P} = \operatorname{diag}(p_1,\dots,p_K)$ with $p_k$ denoting the received power at the $k^\text{th}$ user. 
 For a fixed RF precoder, the only design variables of ZF digital precoder are the received powers, $[p_1,\dots,p_k]$. Using the properties of ZF beamforming; i.e., $|\mathbf{h}_k^H \mathbf{V}_\text{RF} \mathbf{v}_{\text{D}_k}^\text{ZF}|=\sqrt{p_k}$ and $|\mathbf{h}_k^H \mathbf{V}_\text{RF} \mathbf{v}_{\text{D}_\ell}^\text{ZF}|=0$ for all $\ell \not = k$, problem \eqref{main_problem} for designing those powers assuming a feasible RF precoder is reduced to
\begin{subequations}
\label{ZF_digital_problem_MISO}
\begin{eqnarray}
&\displaystyle{\max_{p_1,\dots,p_K \geq 0}}  &   \sum_{k=1}^{K} \beta_k \log_2 \left( 1 + \frac{p_k } {\sigma^2} \right)\\
&\text{s.t.}  & \operatorname{Tr}(\tilde{\mathbf{Q}} \mathbf{P}) \leq P,
\label{ZF_digital_problem_MISO_constraint}
\end{eqnarray}
\end{subequations}
where $\tilde{\mathbf{Q}}= \tilde{\mathbf{V}}_\text{D}^H \mathbf{V}_\text{RF}^H \mathbf{V}_\text{RF} \tilde{\mathbf{V}}_\text{D}$. The optimal solution of this problem can be found by water-filling as
\begin{equation}
\label{water_filling_MISO}
p_k = \frac{1}{\tilde{q}_{kk} }\max\left\{\frac{\beta_k}{\lambda} - \tilde{q}_{kk} \sigma^2,0\right\},
\end{equation}
where ${\tilde{q}}_{kk}$ is $k^\text{th}$ diagonal element of $\tilde{\mathbf{Q}}$ and $\lambda$ is chosen such that $\sum_{k=1}^{K}  \max\{\frac{\beta_k}{\lambda} - \tilde{q}_{kk} \sigma^2,0\}= P$.

\subsection{RF Precoder Design}
Now, we seek to design the RF precoder assuming the ZF digital precoding as in \eqref{ZF_structure}. Our overall strategy is to iterate between the design of ZF precoder and the RF precoder. Observe that the achievable weighted sum rate with ZF precoding in \eqref{ZF_digital_problem_MISO} depends on the RF precoder $\mathbf{V}_\text{RF}$ only through the power constraint \eqref{ZF_digital_problem_MISO_constraint}. Therefore, the RF precoder design problem can be recast as a power minimization problem as
\begin{subequations}
\begin{eqnarray}
&\displaystyle{\min_{\mathbf{V}_\text{RF}}}  &  f(\mathbf{V}_\text{RF})\\
&\text{s.t.} & |\mathbf{V}_\text{RF}(i,j)|^2 = 1, \enspace \forall i,j.
\end{eqnarray}
\end{subequations}
where, $ f(\mathbf{V}_\text{RF})=  \operatorname{Tr}(\mathbf{V}_\text{RF} \tilde{\mathbf{V}}_\text{D}  {\mathbf{P}} \tilde{\mathbf{V}}_\text{D}^H  \mathbf{V}_\text{RF}^H )$.

\begin{algorithm}[t]
\caption{Design of Hybrid Precoders for MU-MISO systems}
\label{Alg:3}
\textbf{Require:} $\beta_k$, $P$, $\sigma^2$
\begin{algorithmic}[1]
\State Start with a feasible $\mathbf{V}_\text{RF}$ and $\mathbf{P}=\mathbf{I}_K$.
				   	\For{$j= 1 \to N^\text{RF}$}\label{step3}
		\State Calculate $\mathbf{A}_j = {\mathbf{P}}^{-\frac{1}{2}} \mathbf{H} 							{\bar{\mathbf{V}}_\text{RF}^{j}}({\bar{\mathbf{V}}_\text{RF}^{j}})^H {\mathbf{H}}^H 						{\mathbf{P}}^{-\frac{1}{2}}$. 
         		\For{$i= 1 \to N$}
         		\State Find  $\zeta_{ij}^{B} $, $\zeta_{ij}^{D} $, $\eta_{ij}^{B}$,
		$\eta_{ij}^{D}$ as defined in Appendix~\ref{Appendix_f_hat}.
		\State Calculate $\theta_{i,j}^{(1)}$ and $\theta_{i,j}^{(2)}$ according to \eqref{theta_solutions_RF}.
         		\State Find $\theta_{ij}^{\text{opt}}= 
         		{\operatorname{argmin}} \left( {\hat{f}(\theta_{i,j}^{(1)}),\hat{f}(\theta_{i,j}^{(2)})}   \right)$.
         		\State Set $\mathbf{V}_\text{RF}(i,j) = e^{-j\theta_{ij}^{\text{opt}}}$.
         		\EndFor
          \EndFor
         	\State Check convergence of RF precoder. If yes, continue; if not go to Step \ref{step3}.
         	\State Find $\mathbf{P} = \operatorname{diag}[{p_1,\dots,p_k}]$ using water-filling as in \eqref{water_filling_MISO}.
         	\State Check convergence of the overall algorithm. If yes, stop; if not go to Step \ref{step3}.
         	\State Set $\mathbf{V}_\text{D} = \mathbf{V}_\text{RF}^H \mathbf{H}^H (\mathbf{H}  \mathbf{V}_\text{RF} \mathbf{V}_\text{RF}^H    \mathbf{H}^H)^{-1} \mathbf{P}^{\frac{1}{2}}$.
\end{algorithmic}
\end{algorithm}

This problem is still difficult to solve since the expression $f(\mathbf{V}_\text{RF})$ in term of $\mathbf{V}_\text{RF}$ is very complicated. But, using the fact that the RF precoder typically satisfies $\mathbf{V}_\text{RF}^H \mathbf{V}_\text{RF} \approx N \mathbf{I}$ when $N$ is large \cite{el2013spatially}, this can be simplified as
\begin{align}
\label{approx_f}\nonumber
f(\mathbf{V}_\text{RF}) &= \operatorname{Tr}(\mathbf{V}_\text{RF}^H \mathbf{V}_\text{RF} \tilde{\mathbf{V}}_\text{D}  {\mathbf{P}} \tilde{\mathbf{V}}_\text{D}^H   )\\ \nonumber
&\approx N  \operatorname{Tr}( {\mathbf{P}}^{\frac{1}{2}}\tilde{\mathbf{V}}_\text{D}^H  \tilde{\mathbf{V}}_\text{D}  {\mathbf{P}}^{\frac{1}{2}}) \\ 
 &= N \operatorname{Tr}\left( (\tilde{\mathbf{H}} \mathbf{V}_\text{RF} \mathbf{V}_\text{RF}^H \tilde{\mathbf{H}}^H )^{-1} \right) = \hat{f}(\mathbf{V}_\text{RF}),
\end{align}
where $\tilde{\mathbf{H}} = {\mathbf{P}}^{-\frac{1}{2}} \mathbf{H}$. Now, analogous to the procedure for the point-to-point	 MIMO case, we aim to extract the contribution of $\mathbf{V}_\text{RF}(i,j)$ in the objective function (here the approximation of the objective function),  $\hat{f}(\mathbf{V}_\text{RF})$, then seek to find the optimal value of $\mathbf{V}_\text{RF}(i,j)$ assuming all other elements are fixed. For $N^\text{RF} > N_s$,
it is shown in Appendix~\ref{Appendix_f_hat} that
\begin{equation}
\label{pure_f_hat}
\hat{f}(\mathbf{V}_\text{RF})  ={N}\operatorname{Tr}(\mathbf{A}_j^{-1}) - {N} \frac{ \zeta_{ij}^{B} +2\operatorname{Re}\bigl\{ {{\mathbf{V}_\text{RF}^{*}}(i,j)} \eta_{ij}^{B}\bigr\}}{ 1+ \zeta_{ij}^{D} +2\operatorname{Re}\bigl\{ {{\mathbf{V}_\text{RF}^{*}}(i,j)} \eta_{ij}^{D}\bigr\} },
\end{equation}
where $\mathbf{A}_j$, $\zeta_{ij}^{B} $, $\zeta_{ij}^{D} $, $\eta_{ij}^{B}$ and
$\eta_{ij}^{D}$ are defined as in Appendix~\ref{Appendix_f_hat} and are independent of $\mathbf{V}_\text{RF}(i,j)$. If we assume that all elements of the RF precoder are fixed except $\mathbf{V}_\text{RF}(i,j) = e^{-j\theta_{i,j}}$, the optimal value for $\theta_{i,j}$ should satisfy $\frac{\partial \hat{f}(\mathbf{V}_\text{RF})}{\partial\theta_{i,j}} =0$. Using the results in Appendix~\ref{Appendix_derivative}, it can be seen that it is always the case that only two $\theta_{i,j} \in [0,2\pi)$ satisfy this condition:
\begin{subequations}
\label{theta_solutions_RF}
\begin{align} 
\theta_{i,j}^{(1)} &= -\phi_{i,j} +\operatorname{sin}^{-1}\left (\frac{z_{ij}}{| c_{ij}|} \right),\\ 
\theta_{i,j}^{(2)} &= \pi  -\phi_{i,j} - \operatorname{sin}^{-1}\left (\frac{z_{ij}}{| c_{ij}|} \right) ,
\end{align}
\end{subequations}
where $c_{ij} = (1+ \zeta_{ij}^{D})\eta_{ij}^{B} -  \zeta_{ij}^{B}\eta_{ij}^{D}$, $z_{ij} = \operatorname{Im}\{ 2(\eta_{ij}^{B})^{*}\eta_{ij}^{D} \}$ and
\begin{align}
\phi_{i,j} =
\begin{cases} {\operatorname{sin}}^{-1}(\frac{\operatorname{Im}\{{c_{ij}}\}}{|c_{ij}|}), &\mbox{if } \operatorname{Re}\{{c_{ij}}\} \geq 0,\\
\pi-{\operatorname{sin}}^{-1}(\frac{\operatorname{Im}\{{c_{ij}}\}}{|c_{ij}|}), & \mbox{if } \operatorname{Re}\{{c_{ij}}\} < 0. \end{cases}
\end{align}

Since $\hat{f}(\mathbf{V}_\text{RF})$ is periodic over $\theta_{i,j}$, only one of those solutions is the minimizer of $\hat{f}(\mathbf{V}_\text{RF})$. The optimal $\theta_{i,j}$ can be written as 
\begin{equation}
\label{arg_max_eq}
\theta_{ij}^{\text{opt}}= \underset{\theta_{i,j}^{(1)},\theta_{i,j}^{(2)}}{\operatorname{argmin}} \left( {\hat{f}(\theta_{i,j}^{(1)}),\hat{f}(\theta_{i,j}^{(2)})}   \right).
 \end{equation}

Now, we are able to devise an iterative algorithm starting from an initially feasible RF precoder and sequentially updating each entry of RF precoder according to \eqref{arg_max_eq} until the algorithm converges to a local minimizer of $\hat{f}(\mathbf{V}_\text{RF}) $.

The overall algorithm is to iterate between the design of $\mathbf{V}_\text{RF}$ and the design of ${\mathbf{P}}$. First, starting with a feasible $\mathbf{V}_\text{RF}$ and $ \mathbf{P}=\mathbf{I}$, the algorithm seeks to sequentially update the phase of each element of RF precoder according to \eqref{arg_max_eq} until convergence. Then, assuming the current RF precoder, the algorithm finds the optimal power allocation $\mathbf{P}$ using \eqref{water_filling_MISO}. The iteration between these two steps continues until convergence. 
The overall proposed algorithm for designing the hybrid digital and analog precoder to maximize the weighted sum rate in the downlink of a multi-user massive MISO system is summarized in Algorithm~\ref{Alg:3}.

\section{Hybrid Beamforming with Finite Resolution Phase Shifters} 
\label{sec:finite-res}
Finally, we consider the hybrid beamforming design with finite resolution phase shifters for the two scenarios of interest in this paper, the point-to-point large-scale MIMO system and the multi-user MISO system with large arrays at the BS. So far, we assume that infinite resolution phase shifters are available in the hybrid structure, so the elements of RF beamformers can have any arbitrary phase angles.
However, components required for accurate phase control can be expensive \cite{krieger2013dense}. Since the number of phase shifters in hybrid structure is proportional to the number of antennas, infinite resolution phase shifter assumption is not always practical for systems with large antenna array terminals. 
In this section, we consider the impact of 
finite resolution phase shifters with $\mathbf{V}_\text{RF}(i,j) \in \mathcal{F}$ and $\mathbf{W}_\text{RF}(i,j) \in \mathcal{F}$ where $\mathcal{F} = \{ 1,\omega, \omega^2, \dots \omega^{n_\text{PS}-1} \}$ and $\omega = e^{j\frac{2\pi}{n_\text{PS}}}$ and $n_\text{PS}$ is the number of realizable phase angles which is typically $n_\text{PS} = 2^b$, where $b$ is the number of bits in the resolution of phase shifters. 

With finite resolution phase shifters, the general weighted sum rate maximization problem can be written as
\begin{subequations}
\label{main_problem_finite_res}
\begin{eqnarray}
&\displaystyle{\Maximize_{\mathbf{V}_\text{RF},\mathbf{V}_\text{D}\mathbf{W}_\text{RF},\mathbf{W}_\text{D}}}  &  \sum_{k=1}^{K} \beta_k R_k\\
&\text{subject to}  &\operatorname{Tr}(\mathbf{V}_\text{RF} \mathbf{V}_\text{D} \mathbf{V}_\text{D}^{H} \mathbf{V}_\text{RF}^{H}) \leq P\\
\label{6c_finite_res}
&&\mathbf{V}_\text{RF}(i,j) \in \mathcal{F}, \enspace \forall i,j\\
&&\mathbf{W}_{\text{RF}_k}(i,j) \in \mathcal{F}, \enspace \forall i,j,k.
\end{eqnarray}
\end{subequations}

For a set of fixed RF beamformers, the design of digital beamformers is a well-studied problem in the literature. However, the combinatorial nature of optimization over RF beamformers in \eqref{main_problem_finite_res} makes the design of  RF beamformers more challenging.
Theoretically, since the set of feasible RF beamformers are finite, we can exhaustively search over all feasible choices. But, as the number of feasible RF beamfomers is exponential in the number of antennas and the resolution of the phase shifters, this approach is not practical for systems with large number of antennas. 

The other straightforward approach for finding the feasible solution for \eqref{main_problem_finite_res} is to first solve the problem under the infinite resolution phase shifter assumption, then to quantize the elements of the obtained RF beamformers to the nearest points in the set $\mathcal{F}$. However, numerical results suggest that for low resolution phase shifters, this approach is not effective. 
This section aims to show that it is possible to account for the finite resolution phase shifter directly in the optimization procedure to get better performance.

For hybrid beamforming design of a single-user MIMO system with finite resolution phase shifters, Algorithm~\ref{Alg:2} for solving the spectral efficiency maximization problem can be adapted as follows. According to the procedure in Algorithm~\ref{Alg:2}, assuming all of the elements of the RF beamformer are fixed except $\mathbf{V}_\text{RF}(i,j)$, we need to maximize $\operatorname{Re}\bigl\{ {{\mathbf{V}_\text{RF}^{*}}(i,j)} \eta_{ij} \bigr\}$ for designing $\mathbf{V}_\text{RF}(i,j)$. This is equivalent to minimizing the angle between $\mathbf{V}_\text{RF}(i,j)$ and $\eta_{ij}$ on the complex plane. Since $\mathbf{V}_\text{RF}(i,j)$ is constrained to be chosen from the set $\mathcal{F}$, the optimal design is
 \begin{equation}
\mathbf{V}^\text{MIMO}_\text{RF}(i,j)  = \mathcal{Q} \left( \psi(\eta_{ij})\right),
 \end{equation}
 where for a non-zero complex variable $a$, $\psi(a)=\frac{a}{|a|}$ and for $a=0$, $\psi(a)=1$,
and the function $\mathcal{Q}(\cdot)$ quantizes a complex unit-norm variable to the nearest point in the set $\mathcal{F}$. \changet{Assuming that the number of antennas at both ends in the same range, i.e., $M = O(N)$, it can be shown that the complexity of the proposed algorithm is polynomial in the number of antennas, $O(N^3)$, while the complexity of finding the optimal beamformers using exhaustive search method is exponential, $O(N^2 2^{bN})$.}

Similarly, for hybrid beamforming design of a MU-MISO system with finite resolution phase shifters, 
Algorithm~\ref{Alg:3} can likewise be modified as follows. Since the set of feasible phase angles are limited, instead of \eqref{arg_max_eq}, we can find $\mathbf{V}_\text{RF}(i,j)$ in each iteration by minimizing $\hat{f}(\mathbf{V}_\text{RF})$ in \eqref{pure_f_hat} using one-dimensional exhaustive search over the set $\mathcal{F}$, i.e., 
\begin{equation}
\mathbf{V}^\text{MU-MISO}_\text{RF}(i,j)  = \underset{\mathbf{V}_\text{RF}(i,j)\in \mathcal{F}}{\operatorname{argmin}} \hat{f}(\mathbf{V}_\text{RF}).
\end{equation}

\changet{The overall complexity of the proposed algorithm for hybrid beamforming design of a MU-MISO system with finite resolution phase shifters is $O(N^2 2^b)$, while the complexity of finding the optimal beamforming using exhaustive search method is $O(N2^{bN})$}. Note that accounting for the effect of phase quantization is most important when low resolution phase shifters are used, i.e., $b=1$ or $b=2$. Since in these cases, the number of possible choices for each element of RF beamformer is small, the proposed one-dimensional exhaustive search approach is not computationally demanding.

\section{Simulations}
In this section, simulation results are presented to show the performance of the proposed algorithms for point-to-point MIMO systems and MU-MISO systems and also to compare them with the existing hybrid beamforming designs and the optimal (or nearly-optimal) fully digital schemes.
In the simulations, the propagation environment between each user terminal and the BS is modeled as a geometric channel with $L$ paths \changet{\cite{liang2014low}}. 
Further, we assume uniform linear array antenna configuration. For such an environment, the channel matrix of the $k^\text{th}$ user can be written as
\begin{equation}
\mathbf{H}_k = \sqrt{\frac{NM}{L}} \sum_{\ell=1}^{L} \alpha_k^{\ell} \mathbf{a}_r(\phi_{r_k}^{\ell}) \mathbf{a}_t(\phi_{t_k}^{\ell})^H ,
\end{equation}
where $\alpha_k^\ell \sim \mathcal{CN}(0,1)$ is the complex gain of the $\ell^{th}$ path between the BS and the user $k$, and $\phi_{r_k}^{\ell} \in [0,~2\pi)$ and $\phi_{t_k}^{\ell} \in [0,~2\pi)$. Further, $\mathbf{a}_r(.)$ and $\mathbf{a}_t(.)$ are the antenna array response vectors at the receiver and the transmitter, respectively. In a uniform linear array configuration with $N$ antenna elements, we have 
\begin{equation}
\mathbf{a}(\phi) = \frac{1}{\sqrt{N}} [1,e^{jk\tilde{d}\sin(\phi)}, \dots,e^{jk\tilde{d}(N-1)\sin(\phi)}]^T,
\end{equation}
where $k=\frac{2\pi}{\lambda}$, $\lambda$ is the wavelength and $\tilde{d}$ is the antenna spacing. 

In the following simulations, we consider an environment with $L=15$ scatterers between the BS and each user terminal assuming uniformly random angles of arrival and departure and $\tilde{d}=\frac{\lambda}{2}$.
For each simulation, the average spectral efficiency is plotted versus signal-to-noise-ratio ($\operatorname{SNR}=\frac{P}{\sigma^2}$) over $100$ channel realizations.

\subsection{Performance Analysis of a MIMO System with Hybrid Beamforming}

\begin{figure}[t]
	\centering
	{\includegraphics[width=0.52\textwidth]{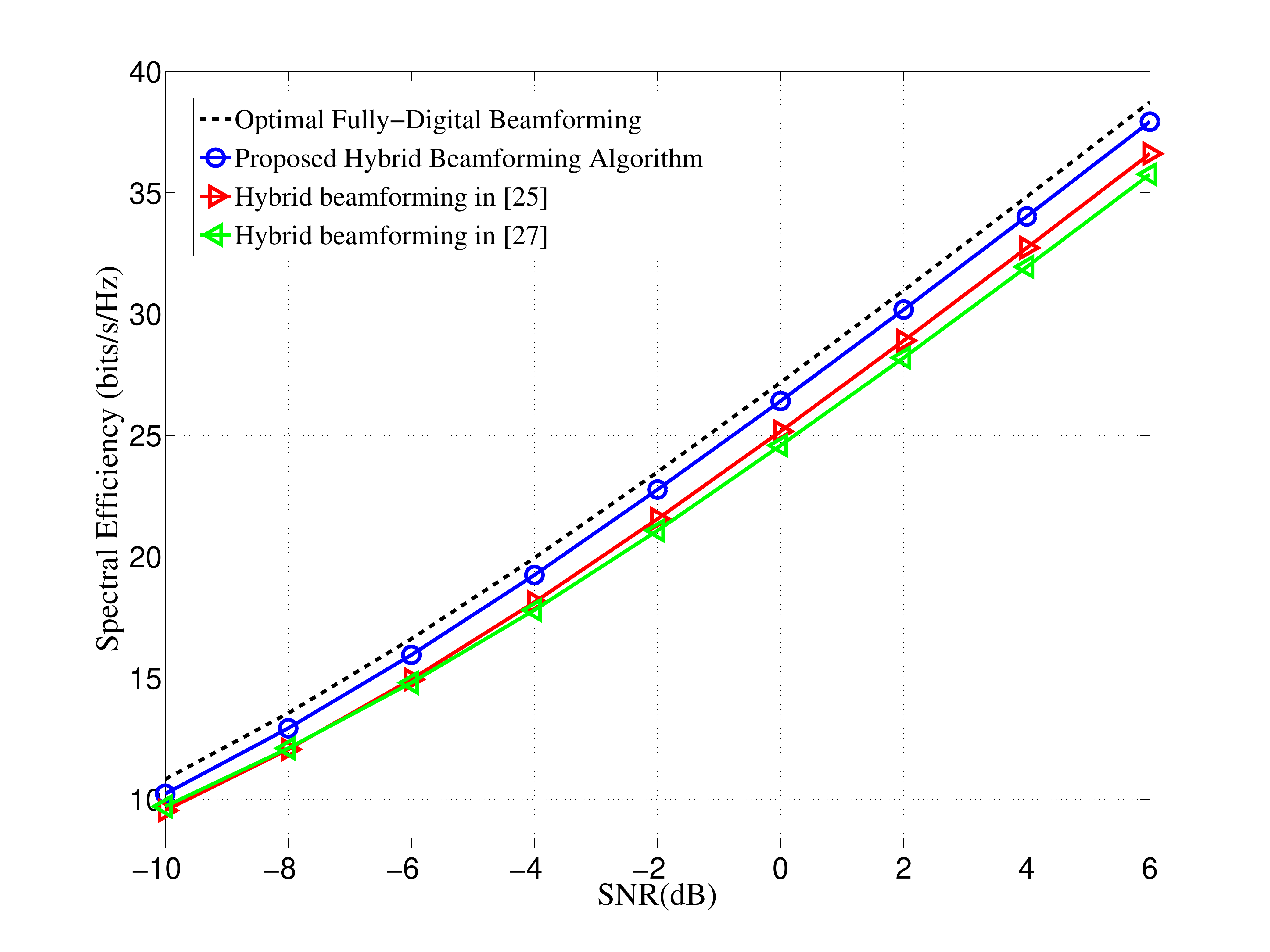}}
	\caption{Spectral efficiencies achieved by different methods in a $64\times16$ MIMO system where $N^\text{RF}=N_s=6$. For hybrid beamforming methods, the use of infinite resolution phase shifters is assumed.}
	\label{fig:simulation_MIMO_infinite}
		\centering
\end{figure}

In the first simulation, we consider a $64\times16$ MIMO system with $N_s = 6$. For hybrid beamforming schemes, we assume that the number of RF chains at each end is $N^\text{RF} = N_s= 6$ and infinite resolution phase shifters are used at both ends. Fig.~\ref{fig:simulation_MIMO_infinite} shows that the proposed algorithm has a better performance as compared to hybrid beamforming algorithms in \cite{el2013spatially} and \cite{zhang2005variable}: about $1.5$dB gain as compared to the algorithm of \cite{el2013spatially} and about $1$dB improvement as compared to the algorithm of \cite{zhang2005variable}. 
Moreover, the performance of the proposed algorithm 
is very close to the rate of optimal fully digital beamforming scheme. This indicates that the proposed algorithm is nearly optimal.

\begin{figure}[t]
	\centering
	{\includegraphics[width=0.52\textwidth]{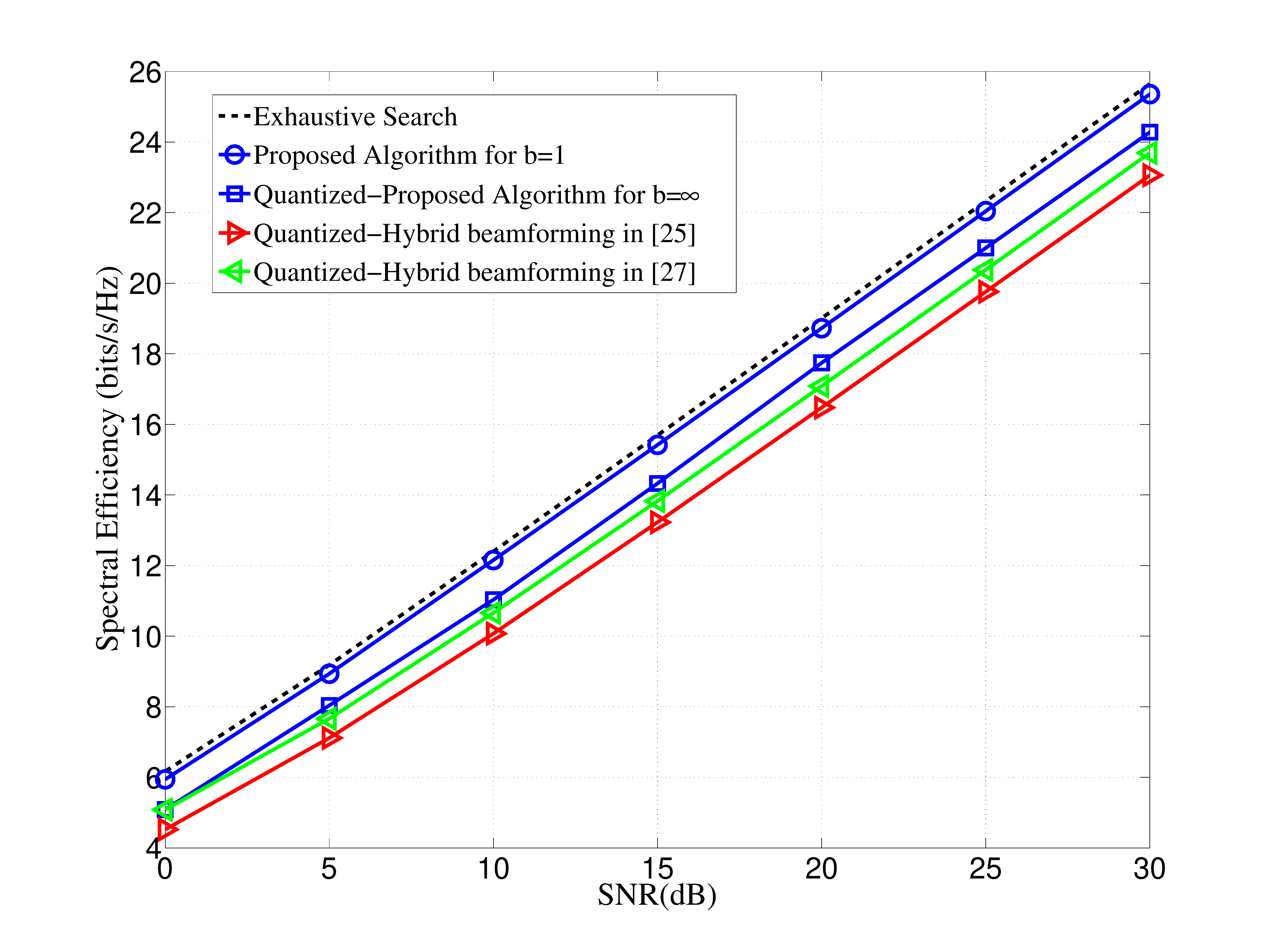}}
	\caption{Spectral efficiencies versus SNR for different methods in a $10\times10$ system where$N^\text{RF}=N_s=2$ and $b=1$.}
	\label{fig:sim1}
\end{figure}

Now, we analyze the performance of our proposed algorithm when only low resolution phase shifters are available. 
First, we consider a relatively small $10\times10$ MIMO system with hybrid beamforming architecture where the RF beamformers are constructed using $1$-bit resolution phase shifters. Further, it is assumed that $N^\text{RF}=N_s=2$. The number of antennas at each end is chosen to be relatively small in order to be able to compare the performance of the proposed algorithm with the exhaustive search method.
We also compare the performance of the proposed algorithm in Section~\ref{sec:finite-res}, which considers the finite resolution phase shifter constraint in the RF beamformer design, to the performance of the quantized version of the algorithms in Section~\ref{sec:MIMO}, and in \cite{zhang2005variable,el2013spatially}, where the RF beamformers are first designed under the assumption of infinite resolution phase shifters, then each entry of the RF beamformers is quantized to the nearest point of the set $\mathcal{F}$. Fig.~\ref{fig:sim1} shows that the performance of the proposed algorithm for $b=1$ has a better performance: at least $1.5$dB gain, as compared to the quantized version of the other algorithms that design the RF beamformers assuming accurate phase shifters first.
Moreover, the spectral efticiency achieved by the proposed algorithm is very close to that of the optimal exhaustive search method, confirming that the proposed methods is near to optimal. 
\begin{figure}[t]
	\centering
	{\includegraphics[width=0.52\textwidth]{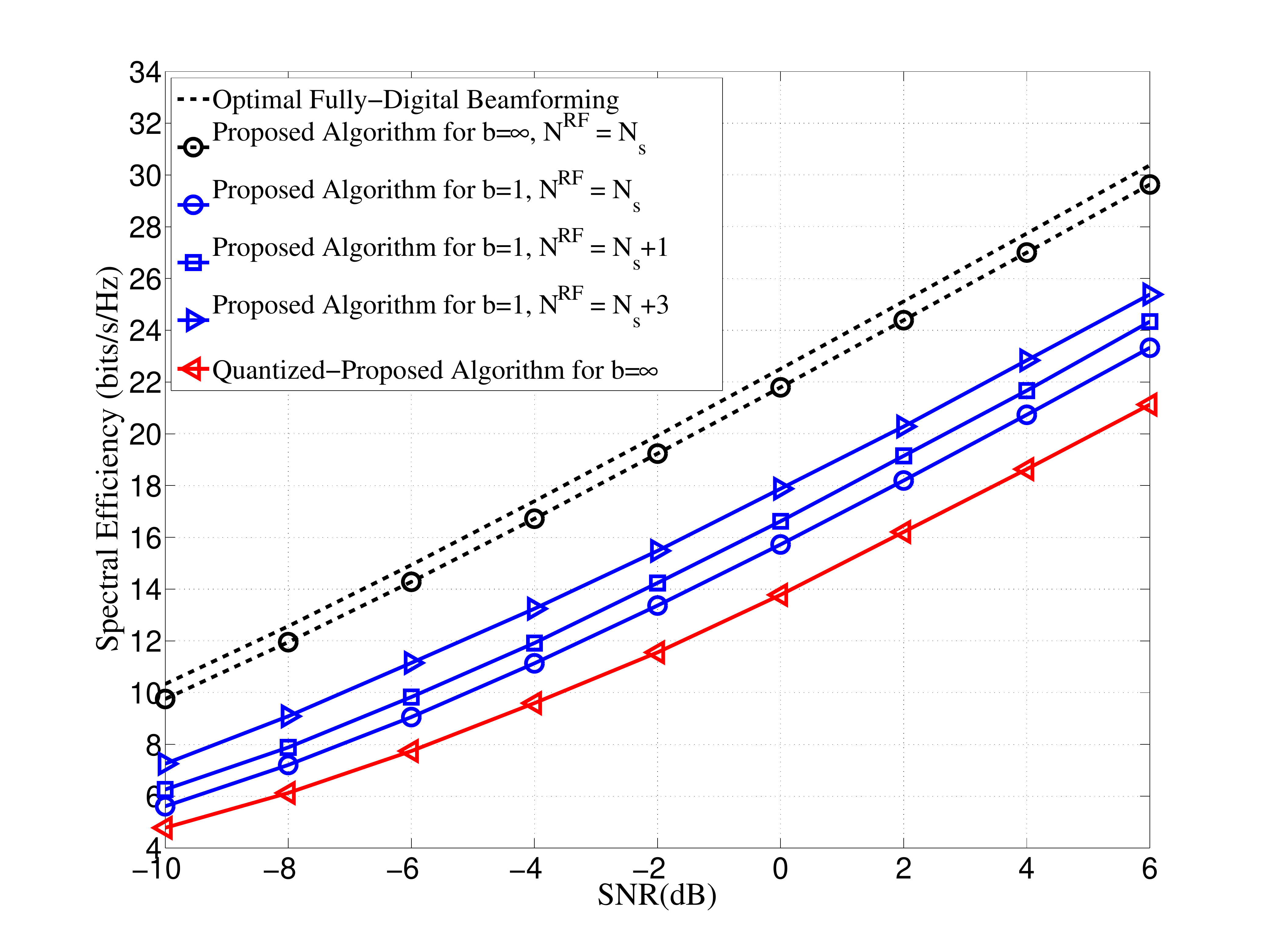}}
	\caption{Spectral efficiencies versus SNR for different methods in a $64\times16$ system where  $N_s=4$.}
	\label{fig:sim2}
\end{figure} 
 
Finally, we consider a $64\times16$ MIMO system with $N_s=4$ to investigate the performance degradation of the hybrid beamforming with low resolution phase shifters. Fig.~\ref{fig:sim2} shows that the performance degradation of a MIMO system with very low resolution phase shifters as compared to the infinite resolution case is significant---about $5$dB in this example. However, Fig.~\ref{fig:sim2} verifies that this gap can be reduced by 
increasing the number of RF chains, 
and by 
using the algorithm in Section~\ref{subsec:between_extremes}
to optimize the RF and digital beamformers. Therefore, the number of RF chains can be used to trade off the accuracy of phase shifters in hybrid beamforming design.

\begin{figure}[t]
	\centering
	{\includegraphics[width=0.52\textwidth]{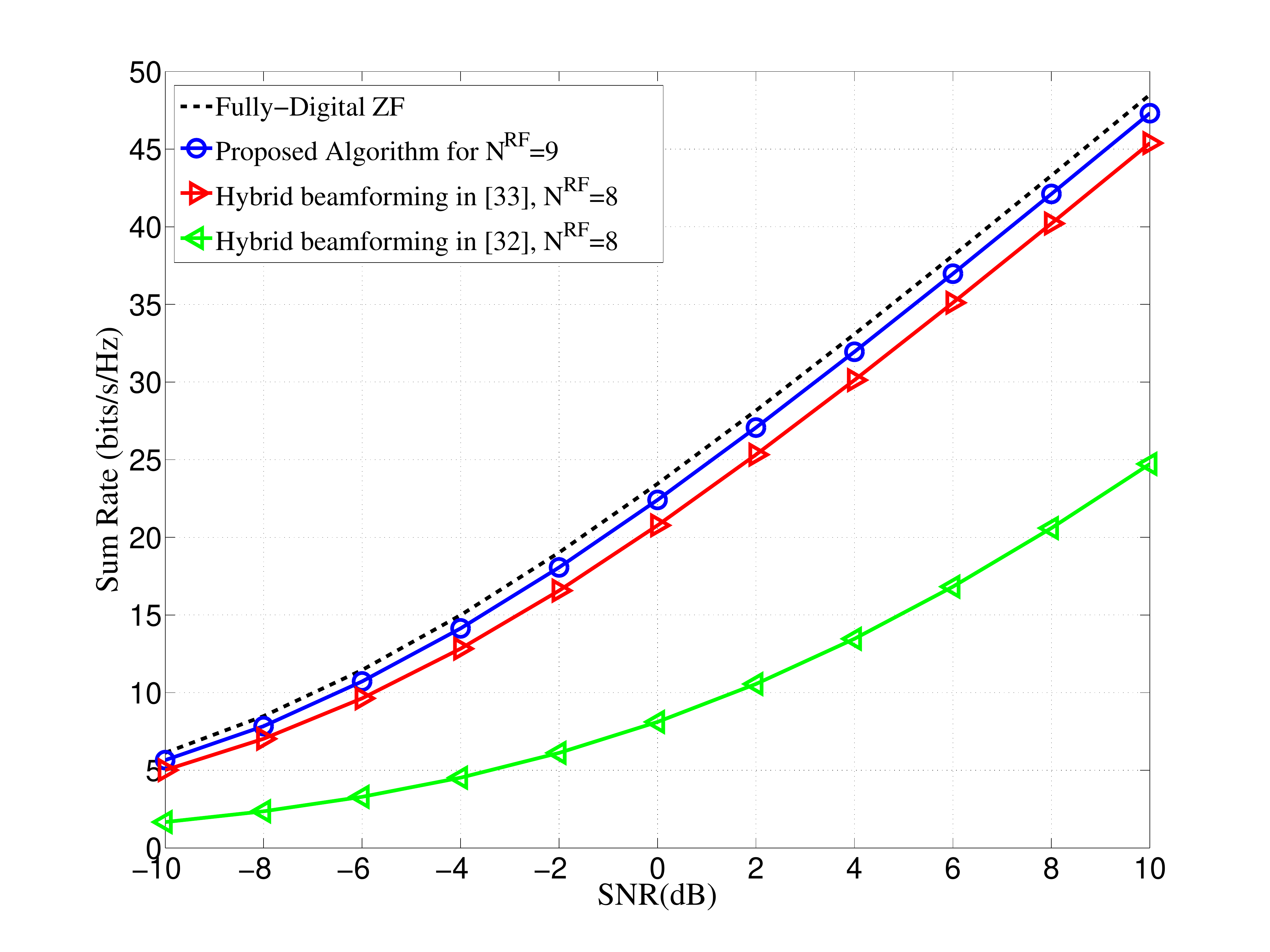}}
	\caption{Sum rate achieved by different methods in an $8$-user MISO system with $N=64$. For hybrid beamforming methods, the use of infinite resolution phase shifters is assumed.}
	\label{fig:simulation_MISO_infinite}
\end{figure}
\begin{figure}[t]
	\centering
	{\includegraphics[width=0.52\textwidth]{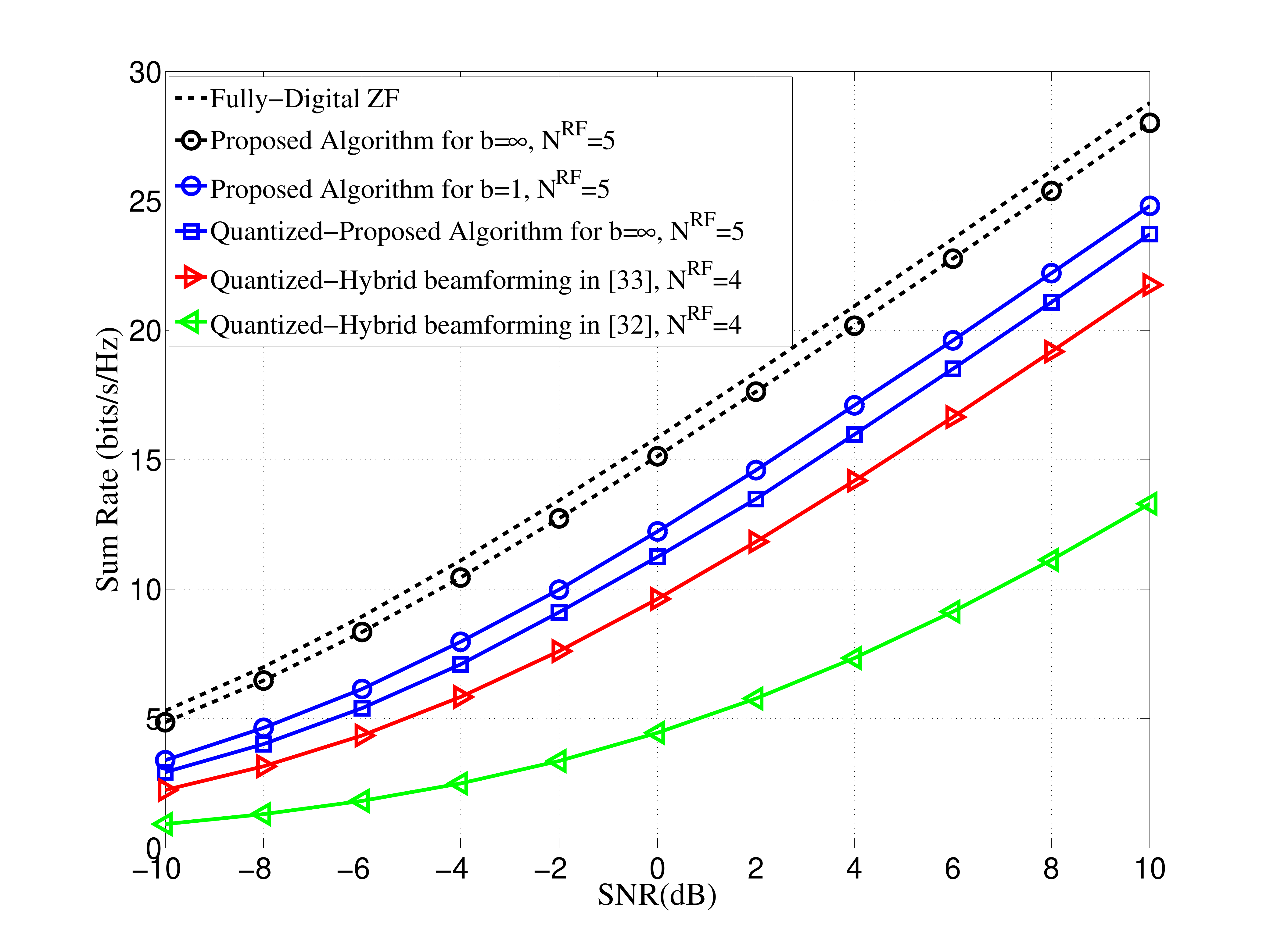}}
	\caption{Sum rate achieved by different methods in a $4$-user MISO system with $N=64$. For the methods with finite resolution phase shifters, $b = 1$.}
	\label{fig:simulation_MISO_finite}
\end{figure}

\subsection{Performance Analysis of a MU-MISO System with Hybrid Beamforming}
To study the performance of the proposed algorithm for MU-MISO systems, we first consider an $8$-user MISO system with $N=64$ antennas at the BS. Further, it is assumed that the users have the same priority, i.e, $\beta_k=1, \forall k$. Assuming the use of infinite resolution phase shifters for hybrid beamforming schemes, we compare the performance of the proposed algorithm with $K+1=9$ RF chains to the algorithms in \cite{liang2014low} and \cite{chinawei} using $K=8$ RF chains. In \cite{liang2014low} and \cite{chinawei} each column of RF precoder is designed by matching to the phase of the channel of each user and matching to the strongest paths of the channel of each user, respectively. Fig.~\ref{fig:simulation_MISO_infinite} shows that the approach of matching to the strongest paths in \cite{chinawei} is not effective for practical value of $N$; (here $N=64$). Moreover, the proposed approach with one extra RF chain
are very close to the sum rate upper bound achieved by fully digital ZF beamforming.
It improves the method in \cite{liang2014low} by about $1$dB in this example.

Finally, we study the effect of finite resolution phase shifters on the performance of the hybrid beamforming in a MU-MISO system. Toward this aim, we consider a MU-MISO system with $N=64$, $K=4$ and $\beta_k=1, \forall k$. Further, it is assumed that only very low resolution phase shifters, i.e., $b=1$, are available at the BS. Fig.~\ref{fig:simulation_MISO_finite} shows that the performance of hybrid beamforming with finite resolution phase shifters can be improved by using the proposed approach
in Section~\ref{sec:finite-res}; it improves the performance about $1$dB, $2$dB and $8$dB respectively as compared to the quantized version of the algorithms in Section~\ref{sec:MIMO}, \cite{liang2014low} and \cite{chinawei} .

\section{Conclusion}
This paper considers the hybrid beamforming architecture for wireless communication systems with large-scale antenna arrays. We show that hybrid beamforming can achieve the same performance of any fully digital beamforming scheme with much fewer number of RF chains; the required number RF chains 
only needs
 to be twice the number of data streams. Further, when the number of RF chains is less than twice the number of data streams, this paper proposes heuristic algorithms for solving the problem of overall spectral efficiency maximization for the
 transmission scenario 
 over a point-to-point MIMO system and over a downlink MU-MISO system. The numerical results show that the proposed approaches achieve a performance close to that of the fully digital beamforming schemes. Finally, we modify the proposed algorithms for systems with finite resolution phase shifters. The numerical results suggest that the proposed modifications can improve the performance significantly, when only very low resolution phase shifters are available. Although the algorithms proposed in this paper all require perfect CSI, they nevertheless serve as benchmark for the maximum achievable rates of the hybrid beamforming architecture.

\appendices
\section{Derivation of \eqref{pure_f_hat}}
\label{Appendix_f_hat}
Let 
$\mathbf{A}_j = \tilde{\mathbf{H}} {\bar{\mathbf{V}}_\text{RF}^{j}} ({\bar{\mathbf{V}}_\text{RF}^{j}})^H \tilde{\mathbf{H}}^H$ 
where 
${\bar{\mathbf{V}}_\text{RF}^{j}}$
  is the sub-matrix of 
  $\mathbf{V}_\text{RF}$ with $j^\text{th}$ column ${\mathbf{v}_\text{RF}^{(j)}}$ removed. 
It is easy to see that 
  $\hat{f}(\mathbf{V}_\text{RF})$ in \eqref{approx_f} can be written as $N \operatorname{Tr}\bigl( (\mathbf{A}_j  + \tilde{\mathbf{H}} \mathbf{v}_\text{RF}^{(j)} {\mathbf{v}_\text{RF}^{(j)}}^H \tilde{\mathbf{H}}^H )^{-1} \bigr)$, 
  where
   $\tilde{\mathbf{H}} \mathbf{v}_\text{RF}^{(j)} {\mathbf{v}_\text{RF}^{(j)}}^H \tilde{\mathbf{H}}^H$
    is a rank one matrix and $\mathbf{A}_j$ is a full-rank matrix for $N^\text{RF} >  N_s$. This enables us to write
\begin{align}
\label{appen_A}
\nonumber
\frac{\hat{f}(\mathbf{V}_\text{RF})}{N} &\equaltext{(a)}  \operatorname{Tr}\left( \mathbf{A}_j^{-1} - \frac{\mathbf{A}_j^{-1}   \tilde{\mathbf{H}} \mathbf{v}_\text{RF}^{(j)} {\mathbf{v}_\text{RF}^{(j)}}^H \tilde{\mathbf{H}}^H \mathbf{A}_j^{-1}  }{1+ \operatorname{Tr}(\mathbf{A}_j^{-1} \tilde{\mathbf{H}} \mathbf{v}_\text{RF}^{(j)} {\mathbf{v}_\text{RF}^{(j)}}^H {\tilde{\mathbf{H}}}^H  )}   \right)\\ \nonumber
 & \equaltext{(b)}  \operatorname{Tr}( \mathbf{A}_j^{-1}) - \frac{ \operatorname{Tr}(  \mathbf{A}_j^{-1}   \tilde{\mathbf{H}} \mathbf{v}_\text{RF}^{(j)} {\mathbf{v}_\text{RF}^{(j)}}^H \tilde{\mathbf{H}}^H \mathbf{A}_j^{-1} ) }{1+ \operatorname{Tr}(\mathbf{A}_j^{-1} \tilde{\mathbf{H}} \mathbf{v}_\text{RF}^{(j)} {\mathbf{v}_\text{RF}^{(j)}}^H \tilde{\mathbf{H}}^H  )}  \\ \nonumber
 & \equaltext{(c)} \operatorname{Tr}( \mathbf{A}_j^{-1}) - 
  \frac{  {\mathbf{v}_\text{RF}^{(j)}}^H \mathbf{B}_j  \mathbf{v}_\text{RF}^{(j)} } 
 {1+ {  {\mathbf{v}_\text{RF}^{(j)}}^H \mathbf{D}_j  \mathbf{v}_\text{RF}^{(j)} }  } \\ 
 & \equaltext{(d)} \operatorname{Tr}(\mathbf{A}_j^{-1}) - \frac{ \zeta_{ij}^{B} +2\operatorname{Re}\bigl\{ {{\mathbf{V}_\text{RF}^{*}}(i,j)} \eta_{ij}^{B}\bigr\}}{ 1+ \zeta_{ij}^{D} +2\operatorname{Re}\bigl\{ {{\mathbf{V}_\text{RF}^{*}}(i,j)} \eta_{ij}^{D}\bigr\} }
\end{align}
where 
\begin{eqnarray*}
 \zeta_{ij}^B &=& \mathbf{B}_j(i,i)\\
 &&+ 2\operatorname{Re} \left \{ \displaystyle{\sum_{m\not=i, n\not=i}{{{\mathbf{V}_\text{RF}^{*}}(m,j)}}\mathbf{B}_j(m,n){\mathbf{V}_\text{RF}(n,j)}} \right \} , \\
\zeta_{ij}^{D} &=& \mathbf{D}_j(i,i) \\
&&+ 2\operatorname{Re} \left \{ \displaystyle{\sum_{m\not=i, n\not=i}{{{\mathbf{V}_\text{RF}^{*}}(m,j)}}\mathbf{D}_j(m,n) {\mathbf{V}_\text{RF}(n,j)} } \right\} , \\
\eta_{ij}^{B} &=& { \sum_{\ell\not=i}} \mathbf{B}_j(i,\ell){\mathbf{V}_\text{RF}(\ell,j)},\\
\eta_{ij}^{D} &=& { \sum_{\ell\not=i}} \mathbf{D}_j(i,\ell) {\mathbf{V}_\text{RF}(\ell,j)},
\end{eqnarray*} 
where $b^j_{i\ell}$ and $d^j_{i\ell}$ 
are the $i^\text{th}$ row and $\ell^\text{th}$ column element of 
$\mathbf{B}_j = \tilde{\mathbf{H}}^H\changet{\mathbf{A}_j^{-2}} \tilde{\mathbf{H}}$ and 
$\mathbf{D}_j = \tilde{\mathbf{H}}^H\changet{\mathbf{A}_j^{-1}}\tilde{\mathbf{H}}$, respectively. In \eqref{appen_A}, the first equality,  $(a)$, is written using the Sherman Morrison formula \cite{bartlett1951inverse}; i.e., $(\mathbf{A}+\mathbf{B})^{-1} = \mathbf{A}^{-1}-\frac{\mathbf{A}^{-1} \mathbf{B} \mathbf{A}^{-1}}{1+\operatorname{Tr(\mathbf{A}^{-1}\mathbf{B})}}$ for a full-rank matrix $\mathbf{A}$ and a rank-one matrix $\mathbf{B}$. Since $\operatorname{Tr}(\cdot)$ is a linear function, equation $(b)$ can be obtained. Equation $(c)$ is based on the fact that the trace is invariant under cyclic permutations\changet{; i.e., $\operatorname{Tr}\left( \mathbf{A} \mathbf{B} \right) =  \operatorname{Tr}\left( \mathbf{B} \mathbf{A} \right)$ for any arbitrary matrices $\mathbf{A}$ and $\mathbf{B}$ with appropriate dimensions}. Finally, $(d)$ is obtained by expanding the terms.  

\section{Derivation of \eqref{theta_solutions_RF}}
\label{Appendix_derivative}
Consider the following function of $\theta$,
\begin{equation}
g(\theta) =  \frac{a_1+2\operatorname{Re}\{b_1 e^{j\theta}\}}{a_2+2\operatorname{Re}\{b_2 e^{j\theta}\}},
=\frac{a_1 + b_1 e^{j\theta} + b_1^{*} e^{-j\theta}}{a_2 + b_2 e^{j\theta} + b_2^{*} e^{-j\theta}}
\end{equation}
where $a_1$ and $a_2$ are real constants and $b_1$ and $b_2$ are complex constants. The maximums and minimums of $g(\theta)$ can be found by solving $\frac{\partial g(\theta)}{\partial \theta} = 0$ or equivalently
\begin{align} \nonumber
\label{append_B2}
&\frac {\partial g(\theta)}{\partial \theta}  = \frac{( j b_1 e^{j\theta} -j b_1^{*} e^{-j\theta})(a_2 + b_2 e^{j\theta} + b_2^{*} e^{-j\theta})}{(a_2 + b_2 e^{j\theta} + b_2^{*} e^{-j\theta})^2}\\
& - \frac{( j b_2 e^{j\theta} -j b_2^{*} e^{-j\theta})(a_1 + b_1 e^{j\theta} + b_1^{*} e^{-j\theta})}{(a_2 + b_2 e^{j\theta} + b_2^{*} e^{-j\theta})^2}  = 0.
\end{align}
By some further algebra, it can be shown that \eqref{append_B2} is equivalent to
\begin{equation}
\label{append_B3}
\operatorname{Im}\{ c e^{j \theta} \} = \operatorname{Im}\{{c}\} \cos(\theta) + \operatorname{Re}\{{c}\}  \sin(\theta)  = z,
\end{equation}
where $z = \operatorname{Im}\{ 2b_1^{*}b_2 \}$ and $c =a_2b_1 - a_1b_2 $. The equation \eqref{append_B3} can be further simplified to
\begin{equation}
\label{append_B4}
|c| \operatorname{sin}(\theta + \phi)  = z,
\end{equation}
where
\begin{align}
\phi =
\begin{cases} {\operatorname{sin}}^{-1}(\frac{\operatorname{Im}\{{c}\}}{|c|}), &\mbox{if } \operatorname{Re}\{{c}\} \geq 0,\\
\pi-{\operatorname{sin}}^{-1}(\frac{\operatorname{Im}\{{c}\}}{|c|}), & \mbox{if } \operatorname{Re}\{{c}\} < 0. \end{cases}
\end{align}
It is easy to show that the \eqref{append_B4} has only two solutions over one period of $2\pi$ as follows:
\begin{subequations}
\begin{align} 
\theta^{(1)} &= -\phi +\operatorname{sin}^{-1}\left (\frac{z}{| c|} \right),\\ 
\theta^{(2)} &= \pi  -\phi - \operatorname{sin}^{-1}\left (\frac{z}{| c|} \right) .
\end{align}
\end{subequations}

\bibliographystyle{IEEEtran}
\bibliography{IEEEabrv,refrences}

\begin{IEEEbiography}[{\includegraphics[width=1in,height=1.25in,clip,keepaspectratio]{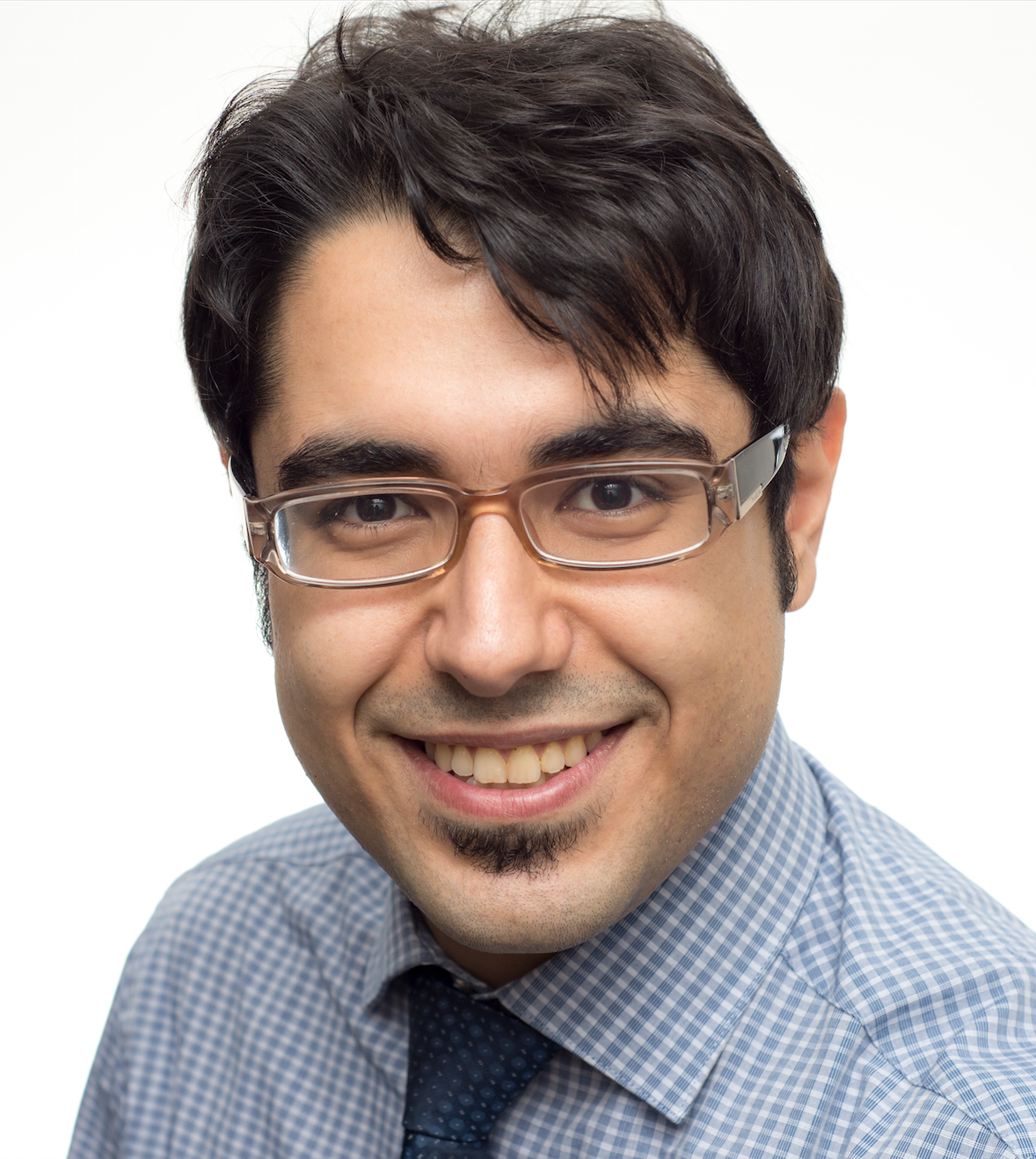}}]{Foad Sohrabi}
(S'13) received his B.A.Sc.\ degree in 2011 from the University of Tehran, Tehran, Iran, and his M.A.Sc.\ degree in 2013 from McMaster University, Hamilton, ON, Canada, both in Electrical and Computer Engineering. Since September 2013, he has been a Ph.D student at University of Toronto, Toronto, ON, Canada. Form July to December 2015, he was a research intern at Bell-Labs, Alcatel-Lucent, in Stuttgart, Germany. His main research interests include MIMO communications, optimization theory, wireless communications, and signal processing.
\end{IEEEbiography}

\begin{IEEEbiography}[{\includegraphics[width=1in,height=1.25in,clip,keepaspectratio]{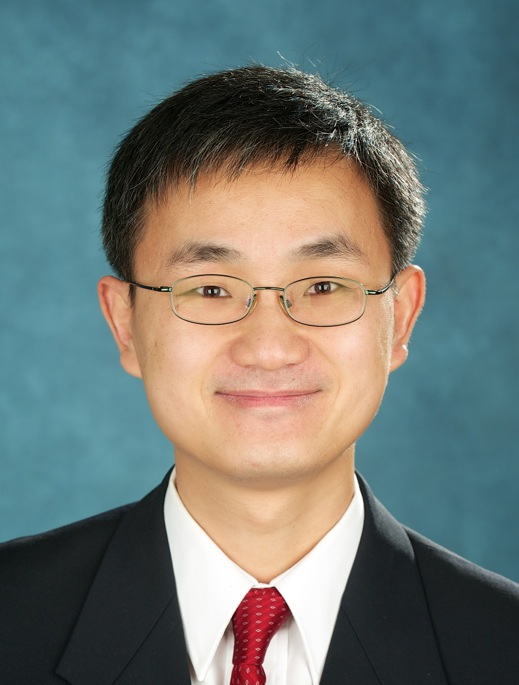}}]{Wei Yu} (S'97-M'02-SM'08-F'14) received the B.A.Sc. degree in Computer Engineering and Mathematics from the University of Waterloo, Waterloo, Ontario, Canada in 1997 and M.S. and Ph.D. degrees in Electrical Engineering from Stanford University, Stanford, CA, in 1998 and 2002, respectively. Since 2002, he has been with the Electrical and Computer Engineering Department at the University of Toronto, Toronto, Ontario, Canada, where he is now Professor and holds a Canada Research Chair (Tier 1) in Information Theory and Wireless Communications. His main research interests include information theory, optimization, wireless communications and broadband access networks.

Prof. Wei Yu currently serves on the IEEE Information Theory Society Board of Governors (2015-17). He is an IEEE Communications Society Distinguished Lecturer (2015-16). He served as an Associate Editor for \textsc{IEEE Transactions on Information Theory} (2010-2013), as an Editor for IEEE \textsc{Transactions on Communications} (2009-2011), as an Editor for \textsc{IEEE Transactions on Wireless Communications} (2004-2007), and as a Guest Editor for a number of special issues for the \textsc{IEEE Journal on Selected Areas in Communications} and the \textsc{EURASIP Journal on Applied Signal Processing}. He was a Technical Program co-chair of the IEEE Communication Theory Workshop in 2014, and a Technical Program Committee co-chair of the Communication Theory Symposium at the IEEE International Conference on Communications (ICC) in 2012. He was a member of the Signal Processing for Communications and Networking Technical Committee of the IEEE Signal Processing Society (2008-2013). Prof. Wei Yu received a Steacie Memorial Fellowship in 2015, an IEEE Communications Society Best Tutorial Paper Award in 2015, an IEEE ICC Best Paper Award in 2013, an IEEE Signal Processing Society Best Paper Award in 2008, the McCharles Prize for Early Career Research Distinction in 2008, the Early Career Teaching Award from the Faculty of Applied Science and Engineering, University of Toronto in 2007, and an Early Researcher Award from Ontario in 2006. He is recognized as a Highly Cited Researcher by Thomson Reuters.
\end{IEEEbiography}

\end{document}